\documentclass[12pt,a4paper]{article}
\usepackage{xcolor}
\usepackage{jheppub}
\usepackage{epstopdf}
\usepackage{graphicx}
\usepackage{epsfig}
\usepackage{dcolumn}  
\usepackage{bm}    
\usepackage{amssymb} 
\usepackage{amsmath,bm}
\usepackage{amsfonts}    
\usepackage{slashed}  
\usepackage{youngtab}
\usepackage{pgf,tikz}
\usepackage{tkz-euclide}
\usetkzobj{all}
\usepackage{pgfplots}

\pgfplotsset{compat=1.8}
\usetikzlibrary{arrows, decorations.pathmorphing, calc,intersections,through,backgrounds,patterns,fit,arrows.meta}

\usetikzlibrary{arrows}
\usepackage[mathscr]{euscript}
\usepackage{epsfig}
\usepackage{pdfpages}
\usepackage{verbatim}
\usepackage{tensor}

\newdimen\tableauside\tableauside=1.0ex
\newdimen\tableaurule\tableaurule=0.4pt
\newdimen\tableaustep
\def\phantomhrule#1{\hbox{\vbox to0pt{\hrule height\tableaurule width#1\vss}}}
\def\phantomvrule#1{\vbox{\hbox to0pt{\vrule width\tableaurule height#1\hss}}}
\def\sqr{\vbox{%
		\phantomhrule\tableaustep
		\hbox{\phantomvrule\tableaustep\kern\tableaustep\phantomvrule\tableaustep}%
		\hbox{\vbox{\phantomhrule\tableauside}\kern-\tableaurule}}}
\def\squares#1{\hbox{\count0=#1\noindent\loop\sqr
		\advance\count0 by-1 \ifnum\count0>0\repeat}}
\def\tableau#1{\vcenter{\offinterlineskip
		\tableaustep=\tableauside\advance\tableaustep by-\tableaurule
		\kern\normallineskip\hbox
		{\kern\normallineskip\vbox
			{\gettableau#1 0 }%
			\kern\normallineskip\kern\tableaurule}%
		\kern\normallineskip\kern\tableaurule}}
\def\gettableau#1 {\ifnum#1=0\let\next=\null\else
	\squares{#1}\let\next=\gettableau\fi\next}

\allowdisplaybreaks[1]

\newcommand{\be}{ \begin{equation}}
\newcommand{\ee}{\end{equation}}
\newcommand{\bea}[1]{\begin{eqnarray}\label{#1} }
\newcommand{\eea}{\end{eqnarray}}

\def\ZZZ{{\hskip-3pt\hbox{ Z\kern-1.6mm Z}}}
\def\zzz{{\hskip-3pt\hbox{ z\kern-1mm z}}}

\newcommand{\KK}{{\cal K}}

\newcommand{\JJ}{{\cal J}}

\newcommand{\abs}[1]{\left| #1 \right|}

\def\one{{\hbox{ 1\kern-.8mm l}}}
\def\zero{{\hbox{ 0\kern-1.5mm 0}}}

\title{A holographic dual for string theory on AdS$_\mathbf{3}\times $S$^\mathbf{3} \times $S$^\mathbf{3} \times $S$^\mathbf{1}$}

\author{Lorenz Eberhardt$^a$, Matthias R.\ Gaberdiel$^a$ and Wei Li$^b$} 
\affiliation{$^a$ Institut f\"ur Theoretische Physik, ETH Zurich, \\
\hspace*{0.3cm}CH-8093 Z\"urich, Switzerland}
\affiliation{$^b$ Institute of Theoretical Physics, Chinese Academy of Science\\
\hspace*{0.3cm} 100190 Beijing, P.R.\ China}
\emailAdd{eberhardtl@itp.phys.ethz.ch, gaberdiel@itp.phys.ethz.ch,  weili@itp.ac.cn}

\abstract{The CFT dual of string theory on $\mathrm{AdS}_3 \times \mathrm{S}^3  \times \mathrm{S}^3 \times \mathrm{S}^1$ is conjectured to be the symmetric orbifold of the $\mathcal{S}_\kappa$ theory, provided that one of the two $Q_5^\pm$ quantum numbers is a multiple of the other. We determine the BPS spectrum of the symmetric orbifold in detail, and show that it reproduces precisely the BPS spectrum that was recently calculated in supergravity. We also determine the BPS spectrum of the world-sheet theory that is formulated in terms of WZW models, and show that, apart from some gaps (which are reminiscent of those that appear in the corresponding $\mathbb{T}^4$ calculation), it also reproduces this BPS spectrum. In fact, the matching seems to work as well as for the familiar $\mathbb{T}^4$ case, and thus our results give strong support for this proposal.}

\begin{document}

\maketitle

\makeatletter
\g@addto@macro\bfseries{\boldmath}
\makeatother
\section{Introduction}
The $\mathrm{AdS}_3/\mathrm{CFT}_2$ duality provides some unique insights into the AdS/CFT correspondence, mainly because of the powerful CFT techniques that are available in two dimensions. 
Indeed, the CFT dual of string theory on $\mathrm{AdS}_3 \times \mathrm{S}^3 \times \mathcal{M}_4$, where $\mathcal{M}_4$ is a 4D hyper-K\"ahler manifold (i.e.\ $\mathbb{T}^4$ or $\mathrm{K3}$), is one of the first explicitly known AdS/CFT dualities constructed from string theory.
On the gravity side, the background arises from the near horizon limit of $Q_1$ D1 branes smeared in $Q_5$ coincident D5 branes that wrap $\mathcal{M}$. 
The dual CFT is then realized as the decoupling limit of the world-volume theory of the D1-D5 system, i.e.\ as a sigma model on the moduli space of $Q_1$ instanton in the $\mathrm{U}(Q_5)$ gauge theory living on $\mathcal{M}_4$. 

As a consequence, the dual CFT lies on the same moduli space as the $Q_1 Q_5$-fold symmetric orbifold of $\mathcal{M}_4$, which is essentially a free theory (at least for the case of $\mathbb{T}^4$). 
Together with the large amount of supersymmetry (described by the small $\mathcal{N}=4$ superconformal algebra) this allows many aspects of the duality to be checked and matched explicitly, see e.g.\ \cite{David:2002wn}  for a review. 
In particular, the BPS spectrum on both sides agrees, as do the three-point functions  \cite{Gaberdiel:2007vu,Dabholkar:2007ey}. 
Finally, for the case of K3, the matching of the elliptic genus provides further non-trivial evidence \cite{deBoer:1998us}. 
\smallskip

For both $\mathbb{T}^4$ and K3, the dual CFT  has small $\mathcal{N}=4$ superconformal symmetry. 
Replacing the hyperk\"ahler manifold $\mathcal{M}_4$ by $\mathrm{S}^3 \times \mathrm{S}^1$ (which is the smallest ``hyperk\"ahler manifold with torsion"), one obtains a close cousin to the above setup, i.e.\ string theory on $\mathrm{AdS}_3 \times \mathrm{S}^3  \times \mathrm{S}^3 \times \mathrm{S}^1$. The dual CFT dual is then expected to have \emph{large} $\mathcal{N}=4$ superconformal symmetry \cite{Elitzur:1998mm,deBoer:1999rh,Gukov:2004ym}. 

Despite its apparent similarity to the familiar $\mathbb{T}^4$ or K3 case, the CFT dual of string theory on $\mathrm{AdS}_3 \times \mathrm{S}^3  \times \mathrm{S}^3 \times \mathrm{S}^1$ has proven much more difficult to find. 
This might be surprising at first sight, given that this background has slightly larger symmetry than the small $\mathcal{N}=4$ algebra supported by the $\mathbb{T}^4$ or K3 case. For example, the CFT dual of the $\mathcal{N}=4$ Vasiliev higher-spin theory was actually first identified for the 
$\mathrm{AdS}_3 \times \mathrm{S}^3  \times \mathrm{S}^3 \times \mathrm{S}^1$ background and shown to be a relatively simple (Wolf) coset CFT \cite{Gaberdiel:2013vva}, while the identification for the $\mathbb{T}^4$ or K3 case was only subsequently found as a limiting case \cite{Gaberdiel:2014cha,Baggio:2015jxa}.

One reason why finding the \emph{stringy} CFT dual of $\mathrm{AdS}_3 \times \mathrm{S}^3  \times \mathrm{S}^3 \times \mathrm{S}^1$ has turned out to be difficult, is that the corresponding brane constructions are rather non-trivial. 
A simple D1-D5 system suffices for the $\mathbb{T}^4$ or K3 case, and the large U-duality symmetry of the system allows one to show that the CFT dual can only depend on the the product $N=Q_1Q_5$ \cite{Seiberg:1999xz}. 
For the $\mathrm{AdS}_3 \times \mathrm{S}^3  \times \mathrm{S}^3 \times \mathrm{S}^1$ case, on the other hand, there are, in addition to the $Q_1$ quantum number, two different $Q_5^\pm$ quantum numbers (corresponding to the sizes of the two ${\rm S}^3$'s). 
We then either need two different kinds of D5 branes, or we must realize one set of charges in terms of some non-trivial flux.
In either case, the description of the moduli space of instantons is much more complicated. 
Furthermore, the U-duality group is much smaller in this case, as was emphasized already in \cite{Gukov:2004ym}; in particular, one should therefore not expect the answer to depend just on some simple combinations of the different brane charges. 

The other main difficulty for finding the CFT dual had to do with the structure of the BPS bounds for the large ${\cal N}=4$ superconformal algebra $A_\gamma$, and its relation to the BPS bound of the corresponding supergravity algebra $D(2,1|\alpha)$. 
In particular, as was stressed in  \cite{deBoer:1999rh, Gukov:2004ym}, the BPS bound for $A_\gamma$ is in general stronger than that for  $D(2,1|\alpha)$, with the bound only coinciding for those BPS states whose spins with respect to the two $\mathfrak{su}(2)$ algebras (corresponding to the two ${\rm S}^3$'s) agree. 
%
Compounding the problem, it was long believed \cite{deBoer:1999rh} that the supergravity theory had lots of BPS states, including many states whose spins with respect to the two $\mathfrak{su}(2)$ algebra do not agree --- and that need to acquire a magical amount of quantum correction in order to just satisfy the $A_\gamma$ BPS bound. Furthermore, none of proposed CFT duals had a corresponding BPS spectrum, even for some special choice of charges \cite{Gukov:2004ym}. 
\smallskip

Recently, this problem was revisited in \cite{Eberhardt:2017fsi}, where it was found that there are no troublesome BPS states (i.e.\ states whose $\mathfrak{su}(2)$-spins do not agree) in supergravity. 
Indeed, motivated by the suggestive results of a world-sheet analysis, we performed a first principle supergravity calculation  \cite{Eberhardt:2017fsi}, and found that the only BPS states that appear in supergravity have the property that their $\mathfrak{su}(2)$ spins agree --- in the old analysis of 
\cite{deBoer:1999rh}, the BPS spectrum of supergravity had only been guessed based on group theoretical methods. 
Furthermore, all the states of supergravity satisfy also the $A_\gamma$ BPS bound --- in fact, this is also true for the non-BPS states in supergravity whose spins differ --- and hence there is no need for any miraculous quantum correction. 




With this roadblock removed, we return in this paper to the search for the CFT dual of string theory on $\mathrm{AdS}_3 \times \mathrm{S}^3  \times \mathrm{S}^3 \times \mathrm{S}^1$.
The dual CFT will be motivated by largely the same methods as those used for the small $\mathcal{N}=4$ case, i.e.\ we start with a D-brane construction and invoke open/closed duality. 
(Proposals of this form were already discussed before, in particular in \cite{Gukov:2004ym}, see also \cite{Elitzur:1998mm}, but they were discarded because of their failure to reproduce the `old' BPS spectrum of supergravity as incorrectly predicted in \cite{deBoer:1999rh}.)
The most promising brane construction appears to be the one where we consider $Q_5^{+}$ D5 branes wrapping   $\mathrm{S}^3 \times \mathrm{S}^1$, where $\mathrm{S}^3$ is the special Lagrangian sub-manifold, supported by $Q_5^{-}$ units of flux, that is wrapped by the $Q_5^+$ D5-branes --- this is the ``third" construction proposed by \cite{Gukov:2004ym}. 
In addition, we add an arbitrary number of $Q_1$ branes smeared on the D5 brane. 
We argue that the dual CFT is then the symmetric orbifold of $\mathrm{S}^3 \times \mathrm{S}^1$, where the flux through the $\mathrm{S}^3$ 
turns out to be $(Q_5^- / Q_5^+) - 1$. 
This can be fairly directly understood for the case when $Q_5^+=1$ since we can then give a direct description of the instanton moduli space; the result for $Q^+_5>1$ is somewhat more conjectural. 
Since the flux has to be quantized, the proposal only makes sense if $Q^{+}_{5}$ is a factor of $Q^{-}_{5}$. 
We also give a microscopic argument (based on anomaly considerations, following \cite{Witten:1999ds}) for where this condition may come from.
\smallskip


We then subject this proposal to some consistency checks. 
In particular, we show that the BPS spectra match from both sides. 
This requires us to determine the BPS spectrum of the symmetric orbifold of $\mathcal{S}_\kappa$ in detail, completing the analysis of \cite{Gukov:2004ym}.
We also compute the full perturbative BPS spectrum from the worldsheet perspective; in particular, we work out the contribution from the spectrally flowed sectors, extending the analysis of \cite{Eberhardt:2017fsi} where only the unflowed sector was analysed. 
As it turns out, this analysis is quite intricate and the resulting BPS spectrum has the same qualitative structure (including some gaps, see the discussion in \cite{Seiberg:1999xz}) as for the familiar case of $\mathbb{T}^4$ and K3.
%

\medskip


This paper is organized as follows. In Section~\ref{sec:branes} we discuss various brane scenarios and discuss their implications on the dual CFT. 
For the case where only one class of D5-branes is present, we can read off an explicit realization of the dual CFT from this picture as the symmetric orbifold of the $\mathcal{S}_\kappa$ \cite{Sevrin:1988ew, Gukov:2004ym} theory. This is only directly possible for $Q_5^+=1$, but we also speculate how the construction should be generalized to $Q_5^+>1$ in Section~\ref{subsec:Q5>1}. Finally, Appendix~\ref{subsec:Skappa algebra} is then devoted to reviewing this theory. 

Section~\ref{sec:BPS spectrum CFT} contains the calculation of the BPS spectrum of the proposed dual --- the symmetric orbifold of $\mathcal{S}_\kappa$. 
This is done carefully and in detail, since there are a number of subtleties (depending on whether the twist of the twisted sector is even or odd, see 
Sections~\ref{subsec:even twists} and \ref{subsec:odd twist}, respectively) that have to be taken into account. Some of the more technical arguments are explained in Appendix~\ref{app:character}, but the main result is simple and spelled out in \eqref{BPS_spectrum Skappa}. 

The next section is concerned with explaining the BPS spectrum of string theory and supergravity on the background $\mathrm{AdS}_3 \times \mathrm{S}^3 \times \mathrm{S}^3 \times \mathrm{S}^1$. Section~\ref{subsec:supergravity} reviews the supergravity calculation of \cite{Eberhardt:2017fsi}, Section~\ref{subsec:worldsheet} the corresponding worldsheet calculation. Section~\ref{subsec:missing chiral primaries T4} clarifies some issues relating to the missing chiral primaries in the case of $\mathbb{T}^4$, while Section~\ref{subsec:BPS spectrum string theory} finally discusses the full perturbative BPS spectrum of string theory, described by \eqref{S3S1_BPS_spectrum}. (Some technical derivations are described in Appendix~\ref{app:BPS states}.)

Section~\ref{sec:BPS comparison} makes some comparisons between the two sides of the duality. First and foremost, we concentrate on the BPS spectrum (Section~\ref{subsec:supergravity and string theory}), but we offer also some further tests by employing the chiral ring (Section~\ref{subsec:chiral ring}) of an $\mathcal{N}=2$ subalgebra of the large $\mathcal{N}=4$ algebra. We also explain how our proposal leads to the symmetric orbifold of $\mathbb{T}^4$ in the infinite radius limit of one of the two three-spheres. Finally, we conclude in Section~\ref{sec:discussion}. For the convenience of the reader we have also reviewed some of the algebras that appear in Appendices~\ref{app:N=4} and \ref{app:saffine}. Furthermore, we have included the derivation of a non-renormalization theorem we have used in the main text in Appendix~\ref{app:chiral}.

\section{Strings on $\mathrm{AdS}_3 \times \mathrm{S}^3 \times \mathrm{S}^3 \times \mathrm{S}^1$ }\label{sec:branes}

In this section we discuss a brane configuration whose near horizon limit gives rise to  $\mathrm{AdS}_3 \times \mathrm{S}^3 \times \mathrm{S}^3 \times \mathrm{S}^1$. 
Since it is engineered from string theory, we can read off from it various aspects of the holographic dual. 
It also guarantees that the duality is consistent non-perturbatively.






\subsection{D5$^{+}$ brane with D5$^{-}$ flux 
}\label{subsec:special lagrangian}

For generic values of the background charges, \cite{Gukov:2004ym} proposed two brane configurations which reproduce different aspects of the geometry. 
In the following we shall concentrate on the one that is more similar to the cases involving $\mathbb{T}^4$ or $\mathrm{K3}$, and that 
gives rise to a natural proposal for the dual CFT. 
To preserve the right amount of supersymmetry, we consider $Q_5^+$ $\mathrm{D5}$-branes wrapping  $\mathrm{S}^3 \times \mathrm{S}^1$, where the $\mathrm{S}^3$ is a special Lagrangian sub-manifold of the six-dimensional D5 world-volume,
which is supported by the flux $Q_5^-$. This construction 
breaks explicitly the symmetry between $Q_5^+$ and $Q_5^-$. We will adopt the convention that the $\mathrm{S}^3$ is located 
in the directions $678$, the $\mathrm{S}^1$ in the direction $9$, i.e.~that the brane configuration is
\be 
\text{
\begin{tabular}{lcccccccccc}
 & 0 & 1 & 2& 3& 4& 5 & 6 & 7 & 8 & 9 \\
 $Q_5^{+}$ $\mathrm{D}5$ branes & $\times$ & & & & & $\times$ & $\times$ & $\times$ & $\times$ & $\times$ \\
$Q_1^{\phantom{+}}$ $\mathrm{D}1$ branes & $\times$ & & & & & $\times$ & $\sim$ &$\sim$ &$\sim$ &$\sim$\\
$Q_5^{-}$ $\mathrm{D}5$ fluxes & & & & & &  &$\circ$  & $\circ$ & $\circ$ &  \\
\end{tabular}
} \label{brane configuration2}
\ee
where $\times$ denotes the directions in which the brane extends, $\sim$ the directions along which the brane is smeared, and $\circ$ denotes fluxes.
The corresponding supergravity was analysed in \cite{Acharya:2000mu, Schvellinger:2001ib, Gauntlett:2001ur, Maldacena:2001pb}. In particular, the brane configuration gives the near-horizon geometry $\mathrm{AdS}_3 \times \mathrm{S}^3 \times \mathrm{S}^3 \times \mathrm{S}^1$.\footnote{The other proposed brane configuration consists of two orthogonal stacks of D5-branes plus D1-branes along their intersection, whose near horizon geometry is $\mathrm{AdS}_3 \times \mathrm{S}^3 \times \mathrm{S}^3 \times \mathbb{R}$ \cite{Cowdall:1998bu, Boonstra:1998yu, Gauntlett:1998kc}. However, we will not consider it in this paper since intersecting five-branes are poorly understood. In particular, it is very hard to 
determine the infrared fixed point of this theory since the dual CFT does not have any interpretation as the moduli space of instantons,  

} 

The resulting gauge theory on the $\mathrm{D5}$ world-volume has 
$\mathcal{N}=2$ supersymmetry, and the three-dimensional low-energy limit is an $\mathcal{N}=2$ Chern-Simons theory with level $
Q_5^-$ living on the directions 059. The bosonic level is however shifted by integrating out the fermions; because of $\mathcal{N}=2$ SUSY, 
this shift is twice as large as the one given in \cite{Witten:1999ds}, so we get a Chern-Simons theory with gauge group $\mathrm{U}(Q_5^+)$ and level 
$Q_5^--Q_5^+$, see also \cite{Gukov:2004ym}. It was also argued in \cite{Witten:1999ds}
that supersymmetry is broken for $Q_5^+ > Q_5^-$, thus we restrict in the following to the case $Q_5^+\le Q_5^-$; the opposite case can be treated similarly by interchanging the roles of $Q_5^+$ and $Q_5^-$.\footnote{
According to \cite{Maldacena:2001pb}, in order for the near horizon limit not to produce singularities for $Q_5^+ \ne Q_5^-$, we must include additional sources in form of additional five-branes wrapping also the other $\mathrm{S}^3$ in the geometry. We will however in the following largely ignore this point, since then we run again into the same problems as for the other brane construction of \cite{Gukov:2004ym}. Furthermore, since this was argued with the help of supergravity, it seems plausible that the full string theory may resolve these singularities.} 

Before we proceed further we should stress that there 
is one crucial difference of our setup relative to the case of $\mathbb{T}^4$. For $\mathbb{T}^4$ we can use four T-dualities in the directions 6789 to arrive again at the same brane configuration, but with the D5-brane charge $Q_5$ and the D1-brane charge $Q_1$ interchanged. This implies that the dual CFT should be symmetric under interchange of $Q_1$ and $Q_5$, a constraint which is obviously satisfied by the $Q_1Q_5$-fold symmetric product of $\mathbb{T}^4$. Importantly, this T-duality is no longer available for $\mathrm{S}^3 \times \mathrm{S}^1$ and so the dual CFT should not be expected to be symmetric in any permutation of $Q_1$, $Q_5^+$ and $Q_5^-$. In fact, the dual CFT is expected to have large ${\cal N}=4$ superconformal symmetry with levels $(Q_1 Q_5^+,Q_1 Q_5^-)$ \cite{Elitzur:1998mm,deBoer:1999rh,Gukov:2004ym} --- for a brief review of this superconformal algebra see Appendix~\ref{app:N=4} --- and central charge 
\be\label{sugrapred}
c = 6 \, Q_{1} \,  \frac{ Q_5^+\, Q_5^-}{Q_5^+ + Q_5^-} \ .
\ee
In particular, this formula does not exhibit any permutation symmetry (except for the obvious $Q_5^+ \leftrightarrow Q_5^-$ exchange symmetry). 

\subsection{The instanton moduli space} \label{subsec:instanton_moduli}

Given that the brane construction of the previous section gives rise to a gauge theory living on the $\mathrm{D5}$ world-volume, we can now follow the usual logic of the AdS/CFT correspondence \cite{Aharony:1999ti}, and identify the dual CFT with the $1+1$-dimensional low-energy theory living on the intersection of the D1- and D5-brane. 
In particular, the D1-branes can be viewed as instantons in the D5-brane theory \cite{Douglas:1995bn}, living on the transverse direction of the D1-branes in the D5-branes, i.e.~on $\mathrm{S}^3 \times \mathrm{S}^1$. 
Low energy fluctuations are described by fluctuations in this moduli space, and thus we can formally identify the dual CFT with the supersymmetric $\sigma$-model on the moduli space $\mathcal{M}_{Q_1,Q_5^+,Q_5^-}$ of $Q_1$ instantons of $\mathrm{SU}(Q_5^+)$ on $\mathrm{S}^3_{Q_5^--Q_5^+} \times \mathrm{S}^1$. 
Here, we have decoupled the overall $\mathrm{U}(1)$ as usual.
The dimension of this moduli space is
\be 
\mathrm{dim}(\mathcal{M}_{Q_1,Q_5^+,Q_5^-})=4 Q_1 Q_5^+\ , \label{dimension_moduli_space}
\ee
as follows from essentially the same argument as for $\mathbb{T}^4$ or $\mathrm{K3}$. 
Indeed, the dimension can be written as twice the absolute value of the index of a Dirac fermion in the adjoint representation. We can then use the Atiyah-Singer Index theorem
\be 
\mathrm{dim}(\mathcal{M}_{Q_1,Q_5^+,Q_5^-})=2 \abs{\mathrm{ind}(\slashed{\nabla}_\mathbf{adj})}=2\abs{\int_{\mathrm{S}^3 \times \mathrm{S}^1} \mathrm{tr}_\mathbf{adj}(e^{i F}) \widehat{A}(R)}\ , 
\ee
where $F$ is the field strength and $R$ the Riemann curvature of an arbitrary metric on $\mathrm{S}^3 \times \mathrm{S}^1$, while 
$\widehat{A}(R)$ is the corresponding Dirac genus. The main point now is that $\mathrm{S}^3 \times \mathrm{S}^1$ is a Lie group: as a consequence, 
its tangent bundle is parallelizable, and all characteristic classes of the tangent bundle vanish, from which we deduce that 
$\widehat{A}(R)=1$ in cohomology. Thus the expression reduces to the usual one
\begin{align} 
\mathrm{dim}(\mathcal{M}_{Q_1,Q_5^+,Q_5^-})&=2\abs{\int_{\mathrm{S}^3 \times \mathrm{S}^1} \mathrm{tr}_\mathbf{adj}(e^{i F})}=2\abs{p_1(F_\mathbf{adj})[\mathrm{S}^3 \times \mathrm{S}^1]}\\
&=4Q_5^+ \abs{p_1(F_\mathbf{f})[\mathrm{S}^3 \times \mathrm{S}^1]}=4\abs{Q_1}Q_5^+\ ,
\end{align}
where we have used the fact that the Dynkin index of the adjoint representation is $Q_5^+$, whereas in the fundamental representation it reads $\tfrac{1}{2}$. Moreover, we have inserted the definition of the instanton number as the first Pontyagin class in the fundamental representation. In the following 
we shall always assume $Q_1 \ge 0$, i.e.~that there are no anti-instantons. We should also mention that 
since the first Pontryagin class of the tangent bundle of $\mathrm{S}^3 \times \mathrm{S}^1$ vanishes, the D5-branes do not induce an 
effective D1-charge \cite{Bershadsky:1995qy}. Thus the identifications of $Q_1$ and $Q_5^+$ made above do not receive corrections.

This moduli space was considered before in the special case $Q_5^+=2$ by mathematicians \cite{Braam:1989aa}. 
It inherits many nice geometric properties from the underlying manifold $\mathrm{S}^3 \times \mathrm{S}^1$. Indeed, the
Hopf surface $\mathrm{S}^3 \times \mathrm{S}^1$ is a hypercomplex manifold, with Hodge diamond
\be \label{hodge}
\begin{tabular}{ccccc}
&& 1 && \\
&0 & & 1 &\\
0& &0 &&0\\
&1 & & 0 &\\
&& 1 && \\
\end{tabular}\ .
\ee
It is moreover  ``hyperk\"ahler with torsion'' (HKT), and supports a $(4,4)$-structure as described in \cite{Moraru:2010aa}. 
As a consequence, the moduli space also supports a HKT and 
a $(4,4)$-structure.\footnote{Note, however, that the terminology is slightly confusing since a HKT manifold is in general not hyperk\"ahler.} 
According to \cite{deBoer:1999rh}, see also \cite{Michelson:1999zf},
this is the geometric requirement that the $\sigma$-model admits classically a large $\mathcal{N}=4$ superconformal symmetry. This gives a consistency check of the identification of the dual CFT with the supersymmetric $\sigma$-model on $\mathcal{M}_{Q_1,Q_5^+,Q_5^-}$.
Furthermore, the Hopf surface $\mathrm{S}^3 \times \mathrm{S}^1$ along with its secondary versions, which are discrete quotients of it, are the only four-dimensional HKT manifolds.

\subsection{The case $Q_5^+=1$} \label{subsecc:instanton_moduli_Q51}

Let us first focus on the case $Q_5^+=1$, for which an explicit description of the moduli space is available. 
To start with we consider the special case where in addition $Q_1=1$. 
Then the moduli space is four dimensional, and since every moduli space of instantons contains the base-manifold as a factor (describing the position of the instanton), and since this has already the right dimension from \eqref{dimension_moduli_space}, we conclude that in this case
\be 
\mathcal{M}_{1,1,Q_5^-}\cong\mathrm{S}^3_{Q_5^--1} \times \mathrm{S}^1\ .
\ee
For general $Q_1$, the moduli space contains the position of all $Q_1$ instantons and hence is naturally identified with the symmetric orbifold \cite{Vafa:1994tf}
\be 
\mathcal{M}_{Q_1,1,Q_5^-}\cong \textrm{Sym}^{Q_1}(\mathrm{S}^3_{Q_5^--1} \times \mathrm{S}^1)
\ , \label{moduli_space}
\ee
where $\textrm{Sym}^{N}(\mathcal{M})\equiv \mathcal{M}^{\otimes N}/S_N$.
Since both $\mathrm{S}^3$ and $\mathrm{S}^1$ are group manifolds, the relevant conformal field theory should simply be the WZW model associated to 
\be
{\cal S}_\kappa \ : \qquad  \mathfrak{su}(2)^{(1)}_{\kappa+2} \oplus \mathfrak{u}(1)^{(1)}  \ , 
\ee
where the superscripts ``(1)" indicate that these are the ${\cal N}=1$ superconformal affine algebras, and $\kappa$ is to be identified with $\kappa=Q_5^--1$. 
(As explained in Appendix~\ref{app:saffine}, the decoupled bosonic $\mathfrak{su}(2)$ algebra of $\mathfrak{su}(2)^{(1)}_{\kappa+2}$ has level $\kappa$, and this should be identified with the flux $Q_5^- -1$.)
This CFT, which was denoted by $\mathcal{S}_\kappa$ in  \cite{Gukov:2004ym}, will play an important role for the rest of the paper, and we have therefore reviewed its salient features in Appendix~\ref{subsec:Skappa algebra}. 
In particular, the $\mathcal{S}_\kappa$ theory supports a  large $\mathcal{N}=4$ algebra with levels $(Q_5^+,Q_5^-)=(1,Q_5^-)$ \cite{Sevrin:1988ew, Gunaydin:1988re}.

By the same arguments as for the familiar $\mathbb{T}^4$ (or K3) case, these considerations then suggest that the CFT dual of string theory on $\mathrm{AdS}_3 \times \mathrm{S}^3 \times \mathrm{S}^3 \times \mathrm{S}^1$ with $Q_5^+=1$ is on the moduli space of the symmetric orbifold 
\be\label{symorb1}
\textrm{Sym}^{Q_1}( \mathcal{S}_\kappa ) \ , \qquad \hbox{with $\kappa=Q_5^- -1$.} 
\ee
This symmetric orbifold has a large ${\cal N}=4$ superconformal symmetry with levels $(Q_1, Q_1 Q_5^-)$, in agreement with the expectations from supergravity, see the discussion around eq.~(\ref{sugrapred}).

\subsection{The case $Q_5^+>1$}\label{subsec:Q5>1}

For $Q_5^+>1$ the identification of the moduli space of instantons is more complicated, and thus the following is somewhat more speculative. In addition, there is a potential subtlety with the world-volume theory. Recall that the low energy effective action of the world-volume theory on the D5-branes includes a Chern-Simons theory with gauge group $\mathrm{U}(Q_5^+)$ and level $Q_5^--Q_5^+$. 
When going to the near-horizon limit of the geometry, the overall $\mathrm{U}(1)$ is decoupled. 
There is then the very subtle issue of whether we end up with a $\mathrm{SU}(Q_5^+)$ or a $\mathrm{SU}(Q_5^+)/\mathbb{Z}_{Q_5^+}$ gauge theory in the end. 
In the case of $\mathrm{AdS}_5 \times \mathrm{S}^5$, it was shown that the center of the group can be spontaneously broken \cite{Aharony:1998qu}. 
It thus seems natural that also $\mathrm{SU}(Q_5^+)/\mathbb{Z}_{Q_5^+}$ Chern-Simons theory with level $Q_5^--Q_5^+$ should be consistent. 

On the other hand, it was argued in \cite{Witten:1999ds} that this theory is anomalous unless $Q_5^+$ divides $Q_5^-$. (Again, one has to make a careful translation of conventions to this paper, since it is formulated in $\mathcal{N}=1$ language.) We are thus led to conclude that the brane scenario we have engineered is only consistent if $Q_5^+$ divides $Q_5^-$ and thus yields also only a prediction on the dual CFT in this case.\footnote{Intuitively, this condition seems to amount to the condition that the flux can be equally divided among the $Q_5^+$ branes.} This requirement was of course trivially satisfied for $Q_5^+=1$.

Obviously, the moduli space of instantons becomes very complicated in the general case, but it still tells us that the dual CFT should be a supersymmetric $\sigma$-model on a $4Q_1Q_5^+$-dimensional HKT space. There is one very natural candidate, which fulfills all the requirements, namely the symmetric orbifold
\be
\textrm{Sym}^{Q_1 Q^+_5} (\mathrm{S}^3_{Q_5^-/Q_5^+-1} \times \mathrm{S}^1)=\textrm{Sym}^{Q_1 Q^+_5} (\mathcal{S}_{\kappa}) \qquad \textrm{with} \quad \kappa=\frac{Q_5^-}{Q_5^+}-1\ . \label{S kappa general}
\ee
This CFT is only unitary if $Q_5^-$ is divisible by $Q_5^+$, since otherwise the WZW-level would not be an integer. This condition is in accordance with the above argument of the brane picture being anomalous. 

We hence propose that \eqref{S kappa general} lies on the same moduli space as string theory on $\mathrm{AdS}_3 \times \mathrm{S}^3 \times \mathrm{S}^3 \times \mathrm{S}^1$ with $Q_5^-/Q_5^+ \in \mathbb{Z}$. 
This proposal generalizes (\ref{symorb1}), to which it reduces for $Q_5^+=1$. 
The opposite case $Q_5^+/Q_5^- \in \mathbb{Z}$ can again be treated by interchanging the roles of $Q_5^+$ and $Q_5^-$. 
Since $\mathcal{S}_\kappa$ supports the large $\mathcal{N}=4$ algebra with levels $(1,\kappa)$, see Appendix~\ref{subsec:Skappa algebra}, the 
symmetric orbifold (\ref{S kappa general}) has levels $(Q_1Q_5^+, Q_1Q_5^-)$, and central charge, see eq.~(\ref{centSk}), 
\be
c= 6 \, Q_1 \, Q_5^+ \frac{Q_5^-}{Q_5^+} \, \frac{1}{\frac{Q_5^-}{Q_5^+}+1} = \frac{6\,  Q_1 Q_5^+ Q_5^-}{Q_5^+ + Q_5^-} \ , 
\ee
in agreement with the supergravity expectation \cite{Gukov:2004ym}, see eq.~(\ref{sugrapred}). 

We should also mention that this proposal generalizes the one given in \cite{Elitzur:1998mm,Argurio:2000tb} for the case of $Q_5^+=Q_5^-$, where it was argued that the dual CFT lies on the same moduli space as the symmetric orbifold of $\mathcal{S}_0$. 
Note that in this case the condition $Q_5^+/Q_5^- \in \mathbb{Z}$ is automatic. We should also mention that the proposal only depends on the products $Q_1Q_5^+$ and $Q_1Q_5^-$. This is not directly in conflict with the missing T-duality of the theory, since the theory is not symmetric in any permutation of the three D-brane charges. However, it implies that there should be some not yet uncovered duality of the bulk theory which acts by rescaling $(Q_1,Q_5^+,Q_5^-) \to (m Q_1,Q_5^+/m,Q_5^-/m)$. It would be very interesting to see this directly. 

Finally, there is yet another consistency condition: for $Q_5^- \rightarrow \infty$, the size of the $\mathrm{S}^3$ in $\mathrm{S}^3 \times \mathrm{S}^1$ tends to infinity, and as in \cite{Gaberdiel:2014cha}, one should expect to make contact with the CFT dual of string theory on $\mathrm{AdS}_3\times \mathrm{S}^3 \times \mathbb{T}^4$. (Strictly speaking, this should only hold for the zero-momentum sector, but this includes in particular the chiral algebra.)
Since $\mathcal{S}_\kappa$ becomes $\mathbb{T}^4$, i.e.~the theory of $4$ free bosons and fermions, for $\kappa\rightarrow \infty$, it follows that 
\be 
\textrm{Sym}^{Q_1 Q^+_5} (\mathrm{S}^3_{Q_5^-/Q_5^+-1} \times \mathrm{S}^1)\  \overset{Q_5^- \to \infty}{\longrightarrow} \ 
\textrm{Sym}^{Q_1 Q^+_5} (\mathbb{T}^4)  \ .
\ee
This provides a very non-trivial test of the proposal; in fact, this requirement seems hard to satisfy with any other candidate.

Since the brane scenario only works for $Q_5^-$ a multiple of $Q_5^+$, this indicates that the dual CFT will be much more complicated in the other cases and not simply given by a symmetric product orbifold. It remains open whether anything more concrete can be said about the dual theories in this case.

\section{BPS spectrum of symmetric orbifold of $\mathcal{S}_\kappa$} \label{sec:BPS spectrum CFT}

As a first check of our proposal we should compare the BPS spectrum of the two theories. 
In this section, we compute the BPS spectrum of the CFT dual, the $N$-fold symmetric orbifold of $\mathcal{S}_\kappa$, where 
\be
N=Q_1 Q^{+}_5 \qquad \textrm{and} \qquad \kappa=\frac{Q_5^{-}}{Q_5^{+}} -1  \ .
\ee
This was already partially done in \cite{Gukov:2004ym}, but we shall be more explicit below, and, as will become apparent, the situation is somewhat more complicated than outlined there. 
We shall only concentrate on the BPS  spectrum from the single-cycle twist sectors of the symmetric orbifold --- these correspond to the single-particle states in AdS. 

This spectrum is to be compared with the predictions coming from the worldsheet theory in string theory (with pure NS-NS flux), as well as supergravity, using the results of \cite{Eberhardt:2017fsi}. 
For this comparison only the low-lying BPS states (whose conformal dimension does not scale with $N=Q_1 Q^{+}_{5}$) play a role, and we shall therefore also concentrate on these BPS states in our analysis of the symmetric orbifold. 

While the detailed analysis is quite complicated --- it will be described in the rest of this section --- the result is simple: the
low-lying single-cycle BPS states of the symmetric orbifold are given by 
%
%
\be 
\bigoplus_{j=0}^{\frac{c}{12}} \, \, [j,j,u=0]_S \otimes \overline{[j,j,u=0]}_S \ ,\label{BPS_spectrum Skappa}
\ee
where $c$ is the total central charge of the symmetric orbifold. Here and in the following $[j^+,j^-,u]_S$ denotes a BPS representation 
of the large $\mathcal{N}=4$ algebra with $\mathfrak{su}(2)$-spins $j^+$ and $j^-$ and $\mathfrak{u}(1)$-charge $u$, see 
\cite{Eberhardt:2017fsi} and Appendix~\ref{app:bound} for our conventions. 


\subsection{BPS spectrum of $\mathcal{S}_\kappa$} \label{subsec:BPS spectrum Skappa}

First, we summarize the BPS spectrum of a single $\mathcal{S}_\kappa$ theory. Representations of the $\mathcal{N}=4$ algebra are labelled by $(h, j^+, j^-, u)$, see e.g.~\cite{Gunaydin:1988re, Gaberdiel:2013vva}. 
The BPS bound for the large $\mathcal{N}=4$ algebra of such a representation is given by, see eq.~(\ref{A28})
\be 
h_\text{BPS}(j^+,j^-,u)=\frac{k^+j^-+k^-j^++(j^+-j^-)^2+u^2}{k^++k^-}\ , \label{BPS_boundNS}
\ee
where $k^+$ and $k^-$ are the levels of the algebra. 
Importantly, the last two terms in the numerator are suppressed for $k^\pm$ large, so they are invisible in the limit in which both levels become large --- recall that for the symmetric orbifold, the relevant levels are $k^\pm = Q_1 Q_5^\pm$, and hence this will be the case for large $Q_1$. 

In the case of $\mathcal{S}_\kappa$, the representations are labelled by $j^- = 0, \tfrac{1}{2},\ldots,\tfrac{1}{2}\kappa$ and $j^+=0$, and hence satisfy the unitarity bound for representations of the large $\mathcal{N}=4$ algebra \cite{Gunaydin:1988re}. Furthermore, $u$ can take any value, and the conformal weight of the corresponding representation is
\be \label{confdim}
h=\frac{j^-(j^-+1)}{\kappa+2}+\frac{u^2}{\kappa+2}=h_\text{BPS}(0,j^-,u)\ .
\ee
Here we have normalized the $\mathfrak{u}(1)$-current in the canonical manner of the large $\mathcal{N}=4$ algebra, see (\ref{A10}); 
depending on the radius of the free boson, the values for $u$ are then quantized in suitable units. 
Since (\ref{confdim}) saturates the BPS bound, all representations of $\mathcal{S}_\kappa$ are in fact BPS. 
We should note that the conformal weight of BPS representations with $u \ne 0$ is not protected under deformations of the theory, since we can continuously change the compactification radius of the boson and hence the conformal dimension. 

\subsection{Untwisted sector} \label{subsec:untwisted sector}

BPS states from the untwisted sector are simple to determine (even if we drop the constraint that they should be `low-lying'). 
Since $j^+=0$ for all representations of $\mathcal{S}_\kappa$, clearly also any symmetric combination will have $j^+=0$. 
To pick out the  representations which result in BPS states, we note that the BPS bound is convex in the sense that 
\be 
\lambda h_\text{BPS}(0,j^-_1,u_1)+(1-\lambda)h_\text{BPS}(0,j^-_2,u_2) \ge h_\text{BPS}(0,\lambda j^-_1+(1-\lambda)j^-_2,\lambda u_1+(1-\lambda) u_2) \ , 
\ee
where $\lambda \in (0,1)$, and we have equality only if $j^-_1=j^-_2$ and $u_1=u_2$. 
Thus only states that arise upon choosing the same representation in all factors can be BPS. 
Thus, the untwisted BPS spectrum consists simply of the BPS states of $\mathcal{S}_\kappa$, except that all quantum numbers have been multiplied by $N$. 
Since the conformal weight of these states scales with $N$, they are not `low-lying' states --- they have the interpretation of different vacua in the bulk, e.g.~black hole geometries.

There are further BPS states which are constructed by the action of the fermions $\psi_{-1/2}^{++}$ on the ground states of the representations. 
They fill in the gaps between the above BPS states, since there are $N$ such fermions, one for each copy. 
Symmetrization introduces however multiple traces (except for the excitation involving a single fermion, which is part of the ${\cal N}=4$ multiplet, see the comment after eq.~(\ref{largeN4BPS})), and thus these states should not be interpreted as being single-particle states in the bulk. 
We should mention that this spectrum is precisely the BPS spectrum which was inferred from the index in \cite{Gukov:2004fh, Gukov:2004ym}. 
This suggests that the index is not very powerful in our context since it can be accounted for purely in terms of untwisted sector states (as well as states from the maximally twisted sector) that are not relevant for the comparison with supergravity. 
This is somewhat similar to the situation encountered in \cite{Maldacena:1999bp}, where for the case of $\mathbb{T}^4$, the index could be obtained just from the groundstate. 

\subsection{Twisted sectors}

Next we consider the BPS states from the single-cycle twisted sectors; in order to simplify the analysis, we shall concentrate from now on the low-lying BPS states. 
They have necessarily the property that $j^+=j^-$ and $u=0$. 
For, if this is not the case, then replacing $N$ by $MN$ the last two terms in the BPS bound (\ref{BPS_boundNS}) decrease by a factor $M$, and hence the given state does not saturate the BPS bound any longer. 
Because of this fact, we will restrict from now on to the zero-charge sector ($u=0$) of the $\mathcal{S}_\kappa$-theory. 

\subsubsection{Odd twist}\label{subsec:odd twist}

\begin{figure}[t]
\begin{center}
\input{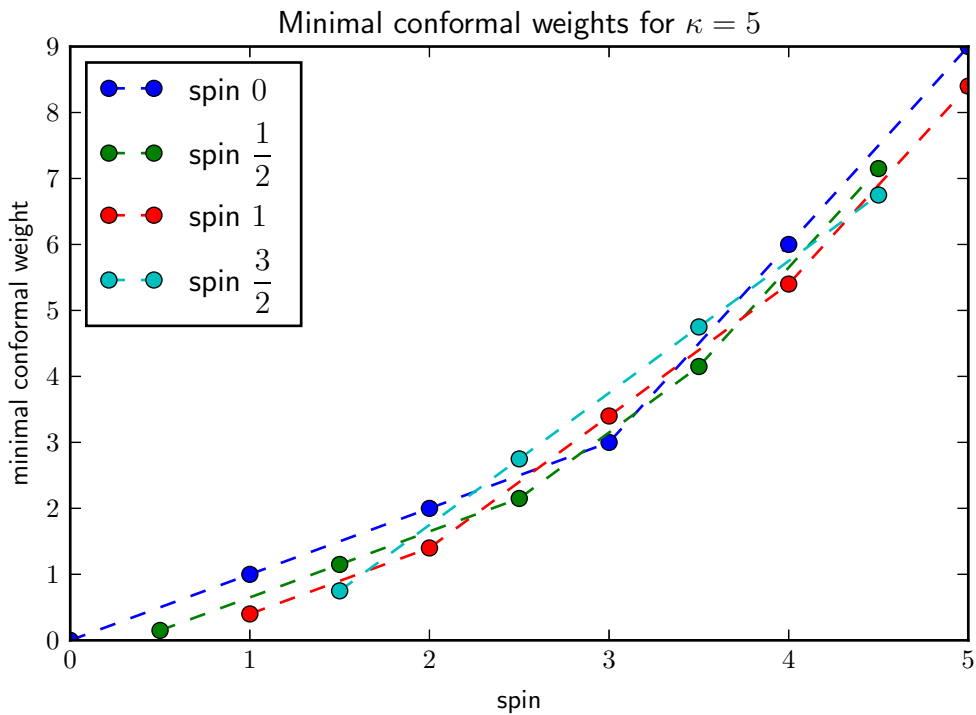}
\end{center}
\caption{Minimal conformal weight for a given spin} \label{fig:su2_minimal_weight}
\end{figure}

The detailed analysis of the twisted sectors depends on whether the twist is odd or even; in this subsection we first deal with the case that the twist $n=2m+1$ is odd --- the even twist case will be discussed in the following subsection. 
Let us recall that a state of weight $h_0$ and charge $R$ (here we collect all possible charges into one index) gives rise to a state in the $n$-twisted sector with charge $R$ and weight 
\be 
h_n=\frac{h_0}{n}+\frac{c(n^2-1)}{24n}\ . \label{twisted_sector_state_odd_eq}
\ee
For a derivation of this fact from the character point of view, see Appendix~\ref{app:character}. 
To obtain minimal conformal weight for a given set of charges, we first need to know the minimal conformal weight at which a given spin of the decoupled $\mathfrak{su}(2)_\kappa$ algebra appears.
This is quite intricate and interesting in its own right 
due to the appearance of null-vectors in the Verma module. 
The situation is depicted in 
Figure~\ref{fig:su2_minimal_weight} for $\kappa=3$, which we have extracted directly from the affine $\mathfrak{su}(2)$-character. 
Clearly, the minimal conformal weight can come potentially from all representations. 
One convenient way to parametrize the minimal conformal weight is as in \cite{Gukov:2004ym}: we introduce a spectral flow of $\mathfrak{su}(2)_\kappa$, which reshuffles states. 
Then the conformal weight and $\mathfrak{su}(2)$-spin is given by
\be 
h_0(j_\mathrm{b})=\frac{j(j+1)}{\kappa+2}+jw+\frac{w^2\kappa}{4}\ , \qquad j_\mathrm{b}^-=j+\frac{1}{2}w \kappa\ . \label{spectral_flow_parametrization}
\ee
This state exists for $j=0,\tfrac{1}{2},\dots,\tfrac{\kappa-1}{2}$ and $w \in \mathbb{Z}_{\ge 0}$. 
Here, we excluded $j=\tfrac{\kappa}{2}$ to avoid overcounting, since, e.g.~$j=\tfrac{\kappa}{2}$, $w=0$ and $j=0$, $w=1$ are the same state.

On top of this state we can apply fractionally moded fermion modes in order to bring the state closer to the BPS bound. 
In terms of the original untwisted state with conformal weight $h_0$ we should apply 
the lowest $m = \frac{n-1}{2}$ $\psi^{++}$ modes  since all of them have conformal weight smaller than $\tfrac{n}{2}$; thus upon dividing by $n$ as in \eqref{twisted_sector_state_odd_eq}, the spin increases more than the conformal weight. 
(Applying $m+1$ of these fermions is also possible ---  this just reflects the fact that each BPS multiplet of the large ${\cal N}=4$ superconformal algebra contains two BPS states that are obtained from one another by the action of the free fermions of the algebra, see the comment after eq.~(\ref{largeN4BPS}).) 
The $m$ $\psi^{++}$ fermions contribute to the conformal weight $\tfrac{1}{2}m^2$. 
We can furthermore also use $p$ fermions of the type $\psi^{+-}$ --- here we adopt the notation that a negative $p$ means $-p$ fermions of the type $\psi^{-+}$ --- but it is easy to see that no other modes can bring one closer to the BPS bound. 
Again, the contribution of these $p$  $\psi^{+-}$ (or $\psi^{-+}$) fermions to the conformal weight is $\tfrac{1}{2}p^2$. 
We then have
\be 
j^+=\frac{1}{2}p+\frac{1}{2}m\ , \quad j^-=j_\mathrm{b}^--\frac{1}{2}p+\frac{1}{2}m \ .
\ee
Using (\ref{twisted_sector_state_odd_eq}) with (\ref{centSk}), the difference to the BPS bound \eqref{BPS_boundNS} is then 
\begin{align} 
\Delta h&=\frac{\kappa+1}{\kappa+2} \frac{m(m+1)}{2m+1}+\frac{m^2+p^2+2h_0(j_\mathrm{b}^-)}{2(2m+1)}\nonumber\\
&\qquad-\frac{(\kappa+1)N(p+m)+N(2j_\mathrm{b}^--p+m)+2(j_\mathrm{b}^--p)^2}{2N(\kappa+2)}  \ .\label{DeltaBPSodd}
\end{align}
We have searched systematically for states saturating this bound, and we have found that for all such states $p=j_\mathrm{b}^-$ and thus $j^+=j^-$, as expected, see the discussion at the beginning of this subsection. 
Then for each $w$ and $j$, there are two solutions of the quadratic equation $\Delta h=0$ in $m$, which are given by
\be 
m=j+\frac{w(\kappa+2)}{2}\ , \quad m=j+\frac{w(\kappa+2)}{2}+\frac{4j}{\kappa}\ . \label{two_solutions_odd}
\ee
Clearly, the second solution is almost always not an integer. There is one exception, however, where we need both solutions, namely for $j=\tfrac{\kappa}{2}$ (which can be identified with a state with $j=0$ and different $w$). 
A convenient way to get both solutions is to relax the above condition that $j \le \tfrac{\kappa-1}{2}$ to $j \le \tfrac{\kappa}{2}$, and only keep the first solution.

We should note that the result has a slightly irregular structure. The associated spin quantum numbers are
\be 
\quad j^+=j^-=j+\frac{1}{2}(\kappa+1)w\ .
\ee
Thus, if $0<j<\tfrac{\kappa-1}{2}$, the solutions are always integer-spaced. However, when changing the spectral flow index $w$, there is one solution which is only half-integer spaced or three-half-integer spaced, depending on whether $\kappa$ is even or odd. As an example, we explicitly list the BPS spectrum for the example $\kappa=5$:
\be 
[0,0]_S\ , \quad [1,1]_S\ , \quad [2,2]_S\ , \quad [\tfrac{7}{2},\tfrac{7}{2}]_S\ , \quad [\tfrac{9}{2},\tfrac{9}{2}]_S\ ,\quad [\tfrac{11}{2},\tfrac{11}{2}]_S\ , \quad [6,6]\ ,\quad \dots
\ee
This irregularity does not occur for $\kappa=0$, since then all representations are half integer spaced. In fact, the second solution of \eqref{two_solutions_odd} is absent in this case. 

\subsubsection{Even twist}\label{subsec:even twists}
Let us now analyse the BPS states that appear in the even twisted sectors, i.e.~for $n=2m$. As explained in Appendix~\ref{app:character}, \eqref{twisted_sector_state_odd_eq} is modified in this case to
\be 
h_n=\frac{h_0}{n}+\frac{cn}{24}+\frac{1}{4n(\kappa+2)}=\frac{h_0}{n}+\frac{c}{24} \frac{n^2-1}{n}+\frac{1}{4n}\ . \label{twisted_sector_state_even}
\ee
In addition, there are fermionic zero modes which generate the representation $(\mathbf{2},\mathbf{1})\oplus(\mathbf{1},\mathbf{2})$, see Appendix~\ref{app:character}. (The additional term $\tfrac{1}{4n}$ 
can be interpreted as coming from the ground state energy of the Ramond fermions \cite{Lunin:2001pw}.) 
Obviously to achieve a BPS state, $j^+$ will be shifted by $+\tfrac{1}{2}$. Furthermore, we can continue to use the parametrization \eqref{spectral_flow_parametrization}, and it is now advantageous to use $m-1$ fermions of the type $\psi^{++}$, and $p$ fermions 
of type $\psi^{+-}$. 
Because of the different moding, they now contribute $\tfrac{1}{2}m(m-1)$ and $\tfrac{1}{2}p(p-1)$ to the conformal weight, respectively. 
Then \eqref{DeltaBPSodd} becomes in this case
\begin{align}
\Delta h&=\frac{\kappa+1}{\kappa+2} \frac{m}{2}+\frac{1}{8m(\kappa+2)}+\frac{m(m-1)+p(p-1)+2h(j_\mathrm{b}^-)}{4m}\nonumber\\
&\qquad-\frac{(\kappa+1)N(p+m)+N(2j_\mathrm{b}^--p+m-1)+2(j_\mathrm{b}^--p-\tfrac{1}{2})^2}{2N(\kappa+2)}\ . \label{DeltaBPSeven}
\end{align} 
Again, an extensive search shows that only states with $p=j_\mathrm{b}^--\tfrac{1}{2}$ and hence $j^+=j^-$ can satisfy the bound. 
There is one exception to this rule, however, namely when $N$ is even: then there are also BPS states in the $N$-twisted sector 
which do not satisfy $j^+=j^-$. But these are high-lying BPS states --- they have to be, given our general argument from the beginning of 
Section~\ref{subsec:odd twist} --- and thus we will not consider them any further for the moment; 
we will come back to them briefly in Section~\ref{sec:BPS comparison}. Similarly, for each given $j$ and $w$, there are two solutions to this equation
\be 
m=j+\frac{w(\kappa+2)}{2}+\frac{1}{2}\ , \quad m=j+\frac{w(\kappa+2)}{2}+\frac{4j}{\kappa}-\frac{1}{2}\ .  \label{two_solutions_even}
\ee
We again relax the restriction $j \le\tfrac{\kappa-1}{2}$ to $j \le \tfrac{\kappa}{2}$ in order to accommodate the cases where the second solution is integer. Luckily, because the terms $\pm \tfrac{1}{2}$ are now present, these conspire in such a way that here the half-integer steps occur where previously 
for odd twists we had three-half-integer steps and vice versa.

Thus, taking together the contributions from the even- and odd-twisted sectors, we obtain a regular BPS spectrum. We illustrate this again for the case $\kappa=5$:
\be 
\begin{tabular}{cccccccccc}
\text{twist} & 1 & 2 & 3 & 4 & 5 & 6 & 8 & 9 &  \\
\text{odd} & $[0,0]_S$ & & $[1,1]_S$ & & $[2,2]_S$ & & & $[\tfrac{7}{2},\tfrac{7}{2}]_S$ & \\
\text{even} & & $[\tfrac{1}{2},\tfrac{1}{2}]_S$ & & $[\tfrac{3}{2},\tfrac{3}{2}]_S$ & & $[\tfrac{5}{2},\tfrac{5}{2}]_S$ & $[3,3]_S$ & &\\
\\
\text{twist}& 10 &  11 & 12 & 13 & 15 & 16 & 17 & 18 & $\cdots$\\
\text{odd}& & $[\tfrac{9}{2},\tfrac{9}{2}]_S$ & & $[\tfrac{11}{2},\tfrac{11}{2}]_S$ & $[6,6]_S$ & & $[7,7]_S$ & & $\cdots$\\
\text{even}& $[4,4]_S$ & & $[5,5]_S$ & & & $[\tfrac{13}{2},\tfrac{13}{2}]_S$ & & $[\tfrac{15}{2},\tfrac{15}{2}]_S$ & $\cdots$
\end{tabular}
\ee
This pattern continues in a similar fashion for generic twists. One can see from this numerical example that there are no BPS states in the twisted sectors of twist
$n=q(\kappa+2)$, where $q \in \mathbb{Z}_{>0}$; we will argue below, see comment (i) in the following subsection, that this holds in general.

\subsection{Recasting and summary of the spectrum}\label{subsec:recasting}

The above method to obtain the BPS spectrum looks quite complicated, and one may be tempted to suspect that there is a simpler
description of the resulting BPS states. Such a formulation can indeed be given by looking at the supersymmetric 
$\mathfrak{su}(2)_1\oplus\mathfrak{su}(2)_{\kappa+1}$-current algebra of $\mathcal{S}_\kappa$. From this perspective, 
the twisted BPS states arise from the $\mathcal{S}_\kappa$ states corresponding to $(0,j)$ upon
applying the spectral flow by $(2j+w(\kappa+1),w)$ with respect to $\mathfrak{su}(2)_1\oplus \mathfrak{su}(2)_{\kappa+1}$. 
Several comments are in order. 
\begin{enumerate}
\item[(i)] From \eqref{two_solutions_odd} and \eqref{two_solutions_even}, the corresponding twist is given by $n=2j+1+w(\kappa+2)$. As promised, we see that $n$ does not attain multiples of $\kappa+2$ and thus these twisted sectors do not contribute BPS states.
\item[(ii)] We see that $j$ and $w$ can be reinterpreted as spins and spectral flow numbers of the $\mathfrak{su}(2)_{\kappa+1}$-algebra,
instead of the bosonic algebra $\mathfrak{su}(2)_\kappa$. This explains why including the case $j=\tfrac{\kappa}{2}$ in the above 
parametrization accounted precisely for all states. Here, this property is natural and manifest.
\item[(iii)] Spectral flow by one unit in one of the $\mathfrak{su}(2)$'s changes the moding of the fermions from NS-moding to R-moding 
and vice versa. Thus the above construction leads to NS-fermions precisely when the sum of both spectral flow parameters is even. 
But from (i) we see that the sum equals precisely the cycle length minus one. 
Thus for odd cycle length NS-fermions occur, while for even cycle length R-fermions occur.
\item[(iv)] The formula for the spectral flow already incorporates the ground state energy of the Ramond fermions as in \eqref{twisted_sector_state_even}. Thus, we should continue to use \eqref{twisted_sector_state_odd_eq} as the ground state energy of the twisted sector.
\item[(v)] Given (ii), (iii) and (iv) above, it is now easy to calculate the conformal weights and spins of the given state to confirm that they are indeed the BPS states we found above. However, it is not very transparent from this perspective why these are the only BPS states --- this is why we 
presented the more pedestrian argument first. 
\end{enumerate}
The BPS spectrum reproduces the one of \cite{Gukov:2004ym}. However, at least on the face of it, the details are a bit different, 
e.g.~in \cite{Gukov:2004ym} the first formulation was used but even and odd twists were treated uniformly. 
We have also shown that all low-lying BPS states have $j^+=j^-$ and $u=0$,
whereas \cite{Gukov:2004ym} only analysed such BPS states (without showing that these are the only ones). 

Let us finally comment on the cutoff of the spectrum. 
It is given by
\be 
j^+_\text{max}=j^-_\text{max}=\frac{N(\kappa+1)}{2(\kappa+2)}=\frac{c}{12}\ , \label{cutoff}
\ee
where $c$ is the central charge of the complete theory. 
In order to see this we note that there are no gaps in the spin spectrum, and that each twisted sector contributes one BPS state, except that there are none for twist $n=q(\kappa+2)$. 
Thus the total number of single-cycle twisted sectors ($N$) needs to multiplied by $\tfrac{\kappa+1}{\kappa+2}$, leading to (\ref{cutoff}). 
The cutoff is of course understood to be the half-integer part of $\tfrac{c}{12}$, i.e.~$\tfrac{1}{2}\lfloor \tfrac{c}{6}\rfloor$. 
In the extreme cases of $\kappa=0$ or $\kappa\to \infty$, we get $\tfrac{1}{4}N$ and $\tfrac{1}{2}N$, respectively. 
The cutoff $\tfrac{1}{2}N$ is indeed familiar from the torus $\mathbb{T}^4$ \cite{Maldacena:1998bw}, which on the level of the vacuum sector can be viewed as the limit of $\mathrm{S}^3 \times \mathrm{S}^1$ with infinite radius, see \cite{Gaberdiel:2014cha}. 
This completes our derivation of eq.~\eqref{BPS_spectrum Skappa}.

\section{BPS spectrum in string theory and supergravity} \label{sec:BPS spectrum string theory supergravity}

In this section we derive the corresponding BPS spectrum in supergravity and in the explicit WZW world-sheet description
of the background.

\subsection{Supergravity} \label{subsec:supergravity}

The BPS spectrum of supergravity on $\mathrm{AdS}_3 \times \mathrm{S}^3 \times \mathrm{S}^3 \times \mathrm{S}^1$
was analysed from first principles in \cite{Eberhardt:2017fsi}, correcting the old analysis of \cite{deBoer:1999rh}. 
It was found there that 
only states with equal spins with respect to the two $\mathfrak{su}(2)$'s are BPS, and that
the whole BPS spectrum organizes itself into representations of the large ${\cal N}=4$ superconformal algebra as 
\be 
\bigoplus_{j\in \frac{1}{2}\mathbb{Z}_{\ge 0}}^\infty [j,j,u=0]_S \otimes \overline{[j,j,u=0]}_S\ . \label{sugra_BPS}
\ee
As before, here $[j,j,u=0]_S$ labels the BPS representation of the $A_\gamma$ algebra, see Appendix~\ref{app:bound} 
for our conventions. In supergravity, only the wedge-modes of the $A_\gamma$ algebra are 
visible, and thus \eqref{sugra_BPS} should be read in the sense of \eqref{largeN4BPS}.

\subsection{World-sheet analysis}\label{subsec:worldsheet}

In the same paper, the BPS spectrum was also determined using the explicit worldsheet description of the background with pure
NS-NS flux in terms of WZW models  \cite{Maldacena:2000hw,Maldacena:2000kv,Maldacena:2001km}, see also 
\cite{Ferreira:2017pgt} and references therein for a description of the supersymmetric setting. 
Since we were only interested in the supergravity limit there, it was sufficient to study only
the unflowed representations, for which the spectrum turned out to be the same as (\ref{sugra_BPS}), but with an upper bound
for the spin $j$ --- this is a consequence of the unitarity bound of \cite{Evans:1998qu}, see also \cite{Hwang:1990aq},
\be 
j\leq \frac{1}{2}\left\lfloor\frac{Q_5^+Q_5^-}{Q_5^++Q_5^-}-1\right\rfloor\ . \label{unitarity_cutoff} 
\ee 
Since $Q_1$ is to be identified with an upper bound on the spectrally flowed sectors ($w\leq Q_1-1$), we need to 
include in general also the spectrally flowed sectors (whose relevance for AdS$_3$ was first recognised in 
\cite{Henningson:1991jc}). The analysis of the BPS spectrum for the spectrally flowed sectors is surprisingly
complicated, and since there are misleading statements in the literature about it,  we shall be fairly 
explicit in the following. The reader who is not interested in these details, may jump directly to Section~5 where
the comparison to the $\mathcal{S}_\kappa$ theories is discussed.
\medskip

In order to explain the spectrally flowed representations in detail, we need to introduce a bit of notation beyond that introduced 
already in \cite{Eberhardt:2017fsi}. The worldsheet theory 
is described in terms of supersymmetric affine theories associated to $\mathfrak{sl}(2,\mathbb{R})_k$ and $\mathfrak{su}(2)_{k^{\pm}}$.
For the spectrally flowed representations, we consider the vector space of the unflowed $\mathfrak{sl}(2,\mathbb{R})_k$
representations, but define on it the action of $\mathfrak{sl}(2,\mathbb{R})_k$ (and the Virasoro algebra) by 
the hatted modes that are related by the spectral flow automorphism to the original modes, see eq.~(\ref{spectralflow}) in Appendix~\ref{app:saffine}. 
With respect to the hatted modes the representation is then not a conventional (Virasoro) highest weight
representation. However, provided that $w>0$, the representation (with respect to the hatted modes)
consists of lowest weight representations of the global $\mathfrak{sl}(2,\mathbb{R})$ algebra --- this
is the case one is interested in, since then the spectrum of the dual CFT will be bounded from below.
In order to describe the resulting spectrum, it is convenient to spectrally flow also in the $\mathfrak{su}(2)_{k^{\pm}}$ 
algebra, although this does not lead to new representations; the relevant spectral flow is also described in 
eq.~(\ref{spectralflow}).

\subsection{Review: full perturbative BPS spectrum for $\mathrm{AdS}_3 \times \mathrm{S}^3 \times \mathbb{T}^4$ }\label{subsec:missing chiral primaries T4}
 
Let us begin by reviewing the more familiar case of $\mathrm{AdS}_3 \times \mathrm{S}^3\times\mathbb{T}^4$. In this case the worldsheet theory consists of an $\mathfrak{sl}(2,\mathbb{R})_k$ WZW model together with a single $\mathfrak{su}(2)_{k'}$ WZW model as well as $4$ free bosons and fermions. 
The requirement that the string theory is critical leads to the condition that $k'=k$, where $k=k'=Q_5$ in the brane description. 

Let us first review the BPS states that come from the unflowed sector. 
In the NS-sector (the analysis for the R-sector is similar) the BPS states arise from the representation with $j_0=j_0'-1$ where $j_0$ and $j_0'$ are the spins of the $\mathfrak{sl}(2,\mathbb{R})$ and $\mathfrak{su}(2)$ ground state representation, respecitvely. 
In each such sector, there are two types of BPS states, ${\cal W}$ corresponding to $j=j_0-1$ and ${\cal Y}$ corresponding to $j=j_0$, see e.g.~\cite{Argurio:2000tb}. 
The ground state spins in the unflowed sector are constrained by the Maldacena-Ooguri bound, which requires that the $\mathfrak{sl}(2,\mathbb{R})$-spin of the 
ground state has to satisfy 
\be 
\frac{1}{2}<j_0 < \frac{k+1}{2}\ . \label{Maldacena-Ooguri bound}
\ee
(Incorporating the continuous representations of $\mathfrak{sl}(2,\mathbb{R})$ amounts to taking the upper bound to be less or equal, so in the following, continuous representations are automatically taken care of.)
In addition to (\ref{Maldacena-Ooguri bound}), there is also the familiar unitarity bound associated to the bosonic $\mathfrak{su}(2)$ algebra at level $k-2$, which requires that 
\be
0\leq j_0' \leq \frac{k-2}{2} \ . 
\ee
Thus from the unflowed sector we get ${\cal W}$- and ${\cal Y}$-type BPS states for 
\begin{equation}\label{unflowed}
\begin{aligned}
{\cal W} : & \quad  j = 0, \ldots ,\tfrac{k-2}{2}  \\
{\cal Y} : & \quad  j = 1 , \ldots, \tfrac{k}{2} \ . 
\end{aligned}
\end{equation}
In order to describe the BPS states in the spectrally flowed sector, we now consider a BPS state in the unflowed sector,  and let it flow simultaneously by the same $w$ in both the $\mathfrak{sl}(2,\mathbb{R})$ and $\mathfrak{su}(2)$ algebra. 
\begin{equation}
\begin{aligned}
L^{\mathfrak{sl}(2,\mathbb{R})}_{0} \, \mapsto \quad &\widehat{L}^{\mathfrak{sl}(2,\mathbb{R})}_{0} = L^{\mathfrak{sl}(2,\mathbb{R})}_{0} - w\, \mathcal{J}_0^3- \frac{k}{4} w^2 \\
L^{\mathfrak{su}(2)}_{0} \, \mapsto \quad&\widehat{L}^{\mathfrak{su}(2)}_{0} \, \, \, =L^{\mathfrak{su}(2)}_{0} \,\,\,+w\, \mathcal{K}_0^3 +\frac{k}{4} w^2 \ . 
\end{aligned}
\end{equation}
Evaluated on the original BPS state for which the eigenvalues of $\mathcal{J}_0^3$ and $\mathcal{K}_0^3$ agree, the total $L_0$ then remains unchanged,
\begin{equation}
L_0=L^{\mathfrak{sl}(2,\mathbb{R})}_{0}+L^{\mathfrak{su}(2)}_{0}  \, \mapsto \quad\widehat{L}_0 = L_0 \ . 
\end{equation}
Thus the flowed state is again on-shell and passes also all the other requirements for being physical. 
Furthermore, since the $\JJ^3_0$ and $\KK^3_0$ eigenvalues are shifted by the same amount (namely $\frac{k}{2}w$), the new state is again BPS, but with a spin that is now 
\begin{equation}
j \mapsto  j + \frac{k}{2}w \ .
\end{equation}

Spectrally flowing thus allows us to obtain BPS states whose spins go beyond the finite range described by (\ref{unflowed}). 
However, the resulting spectrum still has gaps \cite{Argurio:2000tb,Raju:2007uj}. 
In particular, the ${\cal W}$-type BPS states with $j= -\tfrac{1}{2}+\tfrac{k}{2}\mathbb{Z}_{>0}$ do not arise, and for the ${\cal Y}$-type those with 
$j=\tfrac{1}{2}+\tfrac{k}{2}\mathbb{Z}_{>0}$ are missing. 
Combining this with the analysis in the R-sector and putting left- and right-movers together then leads to the BPS spectrum of the worldsheet theory for $\mathbb{T}^4$, given by 
\be 
\bigoplus_{j \in \frac{1}{2}\mathbb{Z}_{\ge 0}\setminus\frac{k}{2}\mathbb{Z}_{>0}} ([j-\tfrac{1}{2}]_S \oplus 2 [j]_S \oplus [j+\tfrac{1}{2}]_S) 
\otimes ( \overline{[j-\tfrac{1}{2}]}_S \oplus 2  \overline{[j]}_S \oplus  \overline{[j+\tfrac{1}{2}]}_S)\ . \label{T4_BPS_spectrum_missing_states}
\ee
Here $[j]_S$ denotes the short representation of the small ${\cal N}=4$ algebra, see, e.g.~\cite{deBoer:1998kjm}.

The missing chiral primaries, i.e.~the missing terms $j \in \tfrac{k}{2}\mathbb{Z}$ break explicitly the T-duality the theory is supposed to have, since the expression is no longer symmetric in $Q_1$ and $Q_5=k$, see the discussion at the end of Section~\ref{subsec:special lagrangian}. 
However, this is believed to be a special feature of the pure NS-NS background, and these missing chiral primaries are expected to be hidden at the instanton singularity \cite{Seiberg:1999xz}. 
The corrected BPS spectrum thus reads
\be 
\bigoplus_{j \in \frac{1}{2}\mathbb{Z}_{\ge 0}}^{\frac{c}{12}} ([j-\tfrac{1}{2}]_S \oplus 2 [j]_S \oplus [j+\tfrac{1}{2}]_S) \otimes 
(\overline{[j-\tfrac{1}{2}]}_S \oplus 2 \overline{[j]}_S \oplus \overline{[j+\tfrac{1}{2}]}_S)\ . \label{T4_BPS_spectrum}
\ee

Finally, in order to understand the upper bound of (\ref{T4_BPS_spectrum}), we note that $Q_1$ should be identified with the maximal winding number as 
\be 
Q_1=\abs{w}+1 \label{Q1_identification}
\ee
since $w+1$ corresponds to the number of fundamental string (one short, and a long one, winding $w$ times around $\mathrm{AdS}_3$). 
Thus the upper bound for $j$ is 
\be
j \le \frac{1}{2} k \, Q_1 = \frac{1}{2}Q_1Q_5=\frac{c}{12} \ . 
\ee

\subsection{The full perturbative BPS spectrum for $\mathrm{AdS}_3 \times \mathrm{S}^3 \times \mathrm{S}^3 \times \mathrm{S}^1$} \label{subsec:BPS spectrum string theory}

For the case of $\mathrm{AdS}_3 \times \mathrm{S}^3 \times \mathrm{S}^3 \times \mathrm{S}^1$, the situation is much more 
complicated because the BPS states in the spectrally flowed sectors do not directly originate from BPS states in the unflowed sector. 
We begin with the technically easier case of the R-sector on the worldsheet. 

\subsubsection{R-sector}\label{app:R-sector}

In the R-sector, the BPS states in the spectrally flowed sectors always originate from ground states before spectral flow. 
Recall from \cite{Eberhardt:2017fsi} that the $L_0$ eigenvalues of the ground states in the unflowed sector are 
\begin{equation}
L^{\mathfrak{sl}(2,\mathbb{R})}_{0} =-\frac{j_0(j_0-1)}{k} \qquad \textrm{and} \qquad L^{\mathfrak{su}(2)^{\pm}}_{0} =\frac{j^{\pm}_0(j^{\pm}_0+1)}{k} \ ,
\end{equation}
where $j_0$ and $j_0^\pm$ denote the $\mathfrak{sl}(2,\mathbb{R})_{k+2}$, $\mathfrak{su}(2)_{k^\pm-2}$ decoupled bosonic spins of the ground state. 
Flowing by $w$, $w^+$ and $w^-$ in $\mathfrak{sl}(2,\mathbb{R})_k$, $\mathfrak{su}(2)_{k^+}$ 
and $\mathfrak{su}(2)_{k^-}$, respectively, the $L_0$ eigenvalues are shifted by
\begin{equation}
\begin{aligned}
L^{\mathfrak{sl}(2,\mathbb{R})}_{0} \, \mapsto \quad &\widehat{L}^{\mathfrak{sl}(2,\mathbb{R})}_{0} = L^{\mathfrak{sl}(2,\mathbb{R})}_{0} - w\, \bigl(j_0-\frac{1}{2}\bigr)- \frac{k}{4} w^2 \\
L^{\mathfrak{su}(2)^{\pm}}_{0} \, \mapsto \quad&\widehat{L}^{\mathfrak{su}(2)^{\pm}}_{0}   =L^{\mathfrak{su}(2)^{\pm}}_{0} +w^{\pm}\, \bigl(j_0^{\pm}+\frac{1}{2}\bigr) +\frac{k}{4}( w^{\pm})^2 \ . 
\end{aligned}
\end{equation}
Note that shift by $\pm \frac{1}{2}$ in the terms proportional to $w$ and $w^\pm$, respectively, comes from the fact that the spectral flow is performed with respect to the full (coupled) algebra, and the spin with respect to the coupled algebra is shifted by $\pm \frac{1}{2}$, see e.g.~\cite{Ferreira:2017pgt} for a detailed discussion. 
(In order to move closer to the BPS bound, we choose the shift so as to lower the $\mathfrak{sl}(2,\mathbb{R})$-spin, but to increase the $\mathfrak{su}(2)$-spins.)
After spectral flow, the resulting spins are then equal to 
\begin{equation}
\begin{aligned}
j=j_0+\frac{kw}{2}-\frac{1}{2} \quad  \textrm{and}\quad  j^\pm=j_0^\pm+\frac{k^\pm w^\pm}{2}+\frac{1}{2}\ . \label{R_sector_spin_relation}
\end{aligned}
\end{equation}
The mass shell condition in the spectrally flowed sector is thus
\begin{align} 
&-\frac{j_0(j_0-1)}{k}-w\left(j_0-\frac{1}{2}\right)-\frac{k}{4}w^2+\frac{j_0^+(j_0^++1)}{k^+}+w^+\left(j^+_0+\frac{1}{2}\right)+\frac{k^+}{4}(w^+)^2\nonumber\\
&\qquad+\frac{j_0^-(j_0^-+1)}{k^-}+w^-\left(j^-_0+\frac{1}{2}\right)+\frac{k^-}{4}(w^-)^2=0\ . \label{R_sector_massshell_condition}
\end{align}
The crucial observation is now that \eqref{R_sector_massshell_condition} can be rewritten in terms of $j$ and $j^\pm$ as
\be 
-\frac{j^2-\tfrac{1}{4}}{k}+\frac{(j^+)^2-\frac{1}{4}}{k^+}+\frac{(j^-)^2-\frac{1}{4}}{k^-}=-\frac{j^2}{k}+\frac{(j^+)^2}{k^+}+\frac{(j^-)^2}{k^-}=0\ . \label{R_sector_massshell_condition_rewritten}
\ee
This gives a relation for $j$ in terms of $j^+$ and $j^-$. 
Combining with the $A_{\gamma}$ BPS bound, one can easily see that there are no BPS states with $j^+\ne j^-$ whose unflowed spin $j_0$ satisfies the Maldacena-Ooguri bound --- this follows by a similar argument as for the unflowed NS sector in \cite{Eberhardt:2017fsi}.
On the other hand, setting
\begin{equation}\label{Rjjpm}
j=j^+=j^-
\end{equation} 
clearly solves equation \eqref{R_sector_massshell_condition_rewritten} since the levels of the coupled algebras must satisfy 
\be\label{kform}
\frac{1}{k} = \frac{1}{k^+} + \frac{1}{k^-} \ , 
\ee
as follows from criticality, see e.g.~\cite{Eberhardt:2017fsi} for more details. 
In addition, the corresponding state is BPS.
Furthermore, by choosing $w$ suitably, any half-integer $j$ can be written in the form \eqref{R_sector_spin_relation} where $j_0$ satisfies the 
Maldacena-Ooguri bound \eqref{Maldacena-Ooguri bound}. Thus, we conclude that there is a BPS state for each half-integer value of $j=j^+=j^-$. 

There is however one subtlety.  Since the spin $j_0^\pm$ of the ground state of the bosonic $\mathfrak{su}(2)_{k^\pm-2}$ algebra must be of the form  $j_0^\pm=0,\frac{1}{2},\dots,\frac{k^\pm-2}{2}$, we can never obtain 
\be
j^{+} \not \in \tfrac{k^+}{2}\mathbb{Z}_{>0} \qquad \textrm{and} \qquad j^{-} \not \in \tfrac{k^-}{2}\mathbb{Z}_{>0} \ . 
\ee
Thus it follows from (\ref{Rjjpm}) that 
\be
j \not \in \tfrac{k^+}{2}\mathbb{Z}_{>0} \cup \tfrac{k^-}{2}\mathbb{Z}_{>0} \ .
\ee
We therefore conclude that BPS states that arise from the R-sector are of the form 
\be 
\bigoplus_{j \in \frac{1}{2}\mathbb{Z}_{> 0} \setminus \left(\frac{k^+}{2}\mathbb{Z}_{> 0}\cup \frac{k^-}{2}\mathbb{Z}_{> 0}\right)} (j,j,u=0)_S\ . \label{BPS_states_R_sector}
\ee
Here, round brackets refer to BPS states, not to complete BPS multiplet. 

Recall from the representation theory of the 
large $\mathcal{N}=4$ superconformal algebra that a BPS multiplet always contains two BPS states, the latter being obtained by 
acting with the fermion $Q^{++}_{-1/2}$ on the highest weight state. 
From the space-time viewpoint, the action of $Q^{++}_{-1/2}$ maps an NS-sector worldsheet state to a R-sector
state and vice versa.
Thus the above states will need to combined with suitable NS-sector states to form full large ${\cal N}=4$ multiplets. 
%
%

\subsubsection{NS sector}
The analysis for the spectrally flowed NS sector is similar, except that in the NS-sector the states before spectral flow are not ground states, but involve also $-\tfrac{1}{2}$ fermion modes. As it turns out, we only need at most one fermionic mode in each of the three algebras, and we can hence parametrize the unflowed state in terms of $\delta,\delta^{\pm}\in \{0,1\}$, where $\delta=0,1$ means that a $\mathfrak{sl}(2,\mathbb{R})$ mode has or has not been applied, and similarly for the two $\mathfrak{su}(2)_{k^\pm}$ algebras. (The relevant mode must again decrease the $\mathfrak{sl}(2,\mathbb{R})$  spin, and increase the 
$\mathfrak{su}(2)$ spins in order to bring the state closer to the BPS bound.) After spectral flow, the massshell condition then reads
\begin{align}
&-\frac{j_0(j_0-1)}{k}-w(j_0-\delta)-\frac{k}{4}w^2+\frac{j_0^+(j_0^++1)}{k^+}+w^+(j_0^++\delta^+)+\frac{k^+}{4}(w^+)^2\nonumber\\
&\qquad+\frac{j_0^-(j_0^-+1)}{k^-}+w^-(j_0^-+\delta^-)+\frac{k^-}{4}(w^-)^2+\frac{1}{2}\delta+\frac{1}{2}\delta^++\frac{1}{2}\delta^-=\frac{1}{2}\ . \label{NS_sector_massshell_condition}
\end{align} 
The true spins $j$, $j^+$ and $j^-$ are given by
\be
j=j_0+\frac{kw}{2}-\delta\ , \quad j^\pm=j_0^\pm+\frac{k^\pm w^\pm}{2}+\delta^\pm\ .
\ee
It is not entirely trivial to find the solutions that saturate the BPS bound, and the detailed construction is described in Appendix~\ref{app:BPS states}. 
%
It follows from the analysis there that the NS-sector BPS spectrum equals\footnote{Strictly speaking, this formula is only true for $k \ge 2$, see the comment below eq.~(\ref{BPS_spectrum_NS_sectorapp}). 
}
\begin{equation}
\begin{aligned} 
&\bigoplus_{j \in \frac{1}{2}\mathbb{Z}_{\ge 0}\setminus\left(\frac{1}{2}\lfloor k \mathbb{Z}_{\ge 0}\rfloor \setminus \frac{1}{2}\,\mathrm{lcm}(k^+,k^-)\mathbb{Z}_{\ge 0}\, \cup \, \left(\frac{1}{2}\lfloor k \mathbb{Z}_{\ge 0}\rfloor+\frac{1}{2}\right) \setminus\left(\frac{1}{2}\,\mathrm{lcm}(k^+,k^-)\mathbb{Z}_{\ge 0}+ \frac{1}{2}\right)\right)} (j,j,u=0)_S \\
&\qquad\qquad\qquad\qquad\ \; \oplus\ \bigoplus_{j \in \frac{k^+}{2}\mathbb{Z}_{> 0} \cup \frac{k^-}{2}\mathbb{Z}_{> 0}} (j,j,u=0)_S\ ,\label{BPS_spectrum_NS_sector}
\end{aligned}
\end{equation}
where as in R sector the round brackets refer to BPS states rather than to complete BPS multiplets. Here $k$ is the level of the supersymmetric $\mathfrak{sl}(2,\mathbb{R})$ WZW model, defined via (\ref{kform}). $\lfloor k \mathbb{Z}_{>0}\rfloor$ denotes the 
elementwise floor of this set --- this is necessary since $k$ is not an integer in the generic case, which complicates the analysis 
significantly. The Maldacena-Ooguri bound \eqref{Maldacena-Ooguri bound} still applies, and it implies that the states 
$\tfrac{1}{2}\lfloor k \mathbb{Z}_{>0} \rfloor$ are located at the boundary between two successive spectrally flowed sectors, 
exactly as in the case of $\mathbb{T}^4$, see \eqref{T4_BPS_spectrum_missing_states}. 

\subsubsection{Full perturbative BPS spectrum}

Combining the BPS states in the R-sector (\ref{BPS_states_R_sector}) with those in the NS-sector (\ref{BPS_spectrum_NS_sector}), we obtain 
\be\label{BPSsRNS}
\bigoplus_{j \in \tfrac{1}{2} \mathbb{Z}_{\ge 0}\setminus \left( \frac{1}{2}\lfloor k \mathbb{Z}_{\ge 0}\rfloor\setminus \frac{1}{2}\, \mathrm{lcm}(k^+,k^-)\mathbb{Z}_{\ge 0}\right)} (j,j,u=0)_S \oplus (j+\tfrac{1}{2},j+\tfrac{1}{2},u=0)_S\ .
\ee
Again, the round brackets refer to BPS states instead of the complete BPS multiplets.
It is very reassuring that (\ref{BPSsRNS}) fits into BPS multiplets (as it has to), i.e.~that the terms in the sum are just the BPS states of the multiplet 
\begin{equation}
[j,j,u=0]_S = (j,j,u=0)_S \oplus (j+\tfrac{1}{2},j+\tfrac{1}{2},u=0)_S
\end{equation}
(that contains two BPS states). 
The above discussion explains also when the R-sector state is the highest weight 
state of the BPS multiplet and when the NS-sector state is. The full spectrum (including right-movers) is then obtained by
tensoring the BPS representations $[j,j,u=0]_S$ (for a given $j$) for left- and right-movers; this then leads to 
%
\be 
\bigoplus_{j \in \frac{1}{2}\mathbb{Z}_{\ge 0} \setminus\left(\frac{1}{2}\lfloor k \mathbb{Z}_{\ge 0} \rfloor 
\setminus \frac{1}{2}\, \mathrm{lcm}(k^+,k^-)\mathbb{Z}_{\ge 0}\right)}^{\frac{c}{12}} [j,j,u=0]_S \otimes \overline{[j,j,u=0]}_S\ . \label{S3S1_BPS_spectrum_missing_states}
\ee
A few comments are in order. First of all, we note that 
\be 
\tfrac{1}{2}\mathbb{Z}_{\ge 0} \cap \tfrac{1}{2} k \mathbb{Z}_{\ge 0} =\tfrac{1}{2}\, \mathrm{lcm}(k^+,k^-) \mathbb{Z}_{\ge 0}\ ,
\ee
so the states $j$, with $j$ divisible by $\tfrac{k^+}{2}$ and $\tfrac{k^-}{2}$ appear in the sum. Secondly, taking the limit of $k^- \to \infty$ in \eqref{S3S1_BPS_spectrum_missing_states} indeed gives back \eqref{T4_BPS_spectrum_missing_states}. Finally, 
the cutoff $\tfrac{c}{12}$ arises in exactly same manner as it did for $\mathbb{T}^4$, i.e.~imposing a maximum 
spectral flow $w$ for $\mathfrak{sl}(2,\mathbb{R})$ by \eqref{Q1_identification} imposes an upper limit on $j$. 

Some of the BPS states can be obtained by spectrally flowing the BPS states of the unflowed sector, as was possible for the case of 
$\mathbb{T}^4$ above. However, in order for this procedure to map BPS states to BPS states, we must now choose the 
spectral flow parameters of the two $\mathfrak{su}(2)$'s as a function of the spectral flow parameter $w$ of $\mathfrak{sl}(2,\mathbb{R})$ as 
\be 
w^+=\frac{k^- w}{k^++k^-}\ , \quad w^-=\frac{k^+ w}{k^++k^-}\ .
\ee
On the other hand, $w^\pm \in \mathbb{Z}$ must be integers, and hence this is not always possible, but only when 
$w$ is a multiple of $(k^++k^-)/\mathrm{gcd}(k^+,k^-)$. Hence this accounts only for a fraction of the BPS states. These special 
states were already identified in \cite{Argurio:2000tb} in the particular case of $k^+=k^-$, but the other BPS states were overlooked. 

Now again by the general logic of \cite{Seiberg:1999xz}, we expect this BPS spectrum to be corrected to
\be 
\bigoplus_{j \in \frac{1}{2}\mathbb{Z}_{\ge 0}}^{\frac{c}{12}} [j,j,u=0]_S \otimes \overline{[j,j,u=0]}_S\ . \label{S3S1_BPS_spectrum}
\ee
This spectrum then agrees with the supergravity spectrum (\ref{sugra_BPS}) in the limit $c\rightarrow \infty$, and this
is the spectrum with which we should compare the BPS spectrum of the dual CFT. 

\section{Comparison between symmetric orbifold and world-sheet} \label{sec:BPS comparison}

After these preparations, it is now straightforward to compare the BPS spectra of the different descriptions. We shall
also make a few additional comments in support of our proposal. 

\subsection{BPS spectrum} \label{subsec:supergravity and string theory}

The low-lying single-particle BPS spectrum of the symmetric orbifold of $\mathcal{S}_\kappa$ was 
calculated in eq.~\eqref{BPS_spectrum Skappa}, and this agrees exactly with the BPS of string theory as 
determined in eq.~\eqref{S3S1_BPS_spectrum}. Apart from the fact that there is a single BPS multiplet for
each half-integer spin, also the upper cutoff matches exactly. This is quite non-trivial, 
since this upper bound arises in quit different ways in the two descriptions. 
\smallskip

\noindent The CFT has three types of additional BPS states, which we now discuss in turn. 
\begin{enumerate}
\item Multi-particle states:  Since the single-particle BPS spectrum agrees on the two sides, and since the states are constructed in exactly the same manner in the CFT and in supergravity or the world-sheet theory, the multi-particle states must also agree.

\item
There are further BPS states arising from the untwisted sector, as discussed in subsection \ref{subsec:untwisted sector}.\footnote{ 
This phenomenon occurs also for $\mathbb{T}^4$ and $\mathrm{K3}$, and it was discussed in detail in \cite{deBoer:1998us} for the case of $\mathrm{K3}$.} 
At conformal weight 
\be 
h=\frac{c}{24}<\frac{c}{12}
\ee
the Ramond ground state appears, which corresponds to a black hole in supergravity, so we expect, a priori, only agreement  of the BPS spectrum up to $\frac{c}{24}$. 
However, as for $\mathbb{T}^4$ and $\mathrm{K3}$, the agreement continues actually up to $\frac{c}{12}$. 
The additional BPS states from the untwisted sector should correspond to BPS states in the geometry $\mathrm{BTZ} \times \mathrm{S}^3 \times \mathrm{S}^3 \times \mathrm{S}^1$. 
The cutoff $\frac{c}{12}$ seems to be a generic feature of string holography on $\mathrm{AdS}_3$-backgrounds.

\item
Finally, the symmetric orbifold of $\mathcal{S}_\kappa$ possesses additional BPS states in the maximally twisted sector, 
provided that the number of copies is even. This corresponds to a deeply stringy effect and hence cannot be seen in supergravity, nor in 
the worldsheet description, where the emergence of $Q_1$ is somewhat hidden. 
The relevant states all have spins $j^+\neq j^-$, and therefore do not give rise to ${\cal N}=2$ chiral primaries, as will be discussed in the following subsection. 
\end{enumerate}

\subsection{The chiral ring} \label{subsec:chiral ring}

The large $\mathcal{N}=4$ superconformal algebra contains an $\mathcal{N}=2$ superconformal algebra as a subalgebra \cite{Gunaydin:1988re, Gukov:2004ym}, see Appendix~\ref{app:N2subalgebra} for a short review. 
It is not difficult to see that the only BPS states of the large $\mathcal{N}=4$ superconformal algebra that are also BPS with respect to this ${\cal N}=2$ subalgebra are those with $j^+=j^-$ and $u=0$. 
Indeed, the BPS bound of the $\mathcal{N}=2$ algebra agrees with the one of $\mathrm{D}(2,1|\alpha)$ \eqref{A30}, as one can see from \eqref{N2subalgebra}. 
This implies that the only $\mathcal{N}=2$ chiral primaries have $j^+=j^-$ and $u=0$, since otherwise the state would violate the $\mathcal{N}=4$ BPS bound. 
The highest weight state with respect to the two $\mathfrak{su}(2)$-algebras is a chiral primary state, while the lowest weight state describes an anti-chiral primary. 
This explains also why the BPS bound \eqref{BPS_boundNS} has such a simple form in this case. 

It is very intriguing that the only BPS multiplets of supergravity are of this form, i.e.~contain ${\cal N}=2$ chiral (or anti-chiral) primary states. 
In particular, it is known that the spectrum of $\mathcal{N}=2$ chiral primaries is invariant under deformations,\footnote{For the 
benefit of the reader, we have recalled the argument building upon \cite{Dixon:1987bg, deBoer:2008ss, Baggio:2012rr} 
in Appendix~\ref{app:chiral} showing that these primaries are stable to all orders of conformal perturbation theory.} 
and thus we should find the same chiral primary spectrum at all points in moduli space. This is nicely borne out by our analysis:
the only additional BPS states of the symmetric orbifold of $\mathcal{S}_\kappa$ (that appear in the maximally twisted sector
provided that the number of copies is even) always have different spins $j^+\neq j^-$, and hence do not give rise to (stable)
${\cal N}=2$ chiral primary states. It also ties in nicely with the conclusions reached in \cite{Baggio:2017kza} where using integrability arguments it was argued that the BPS spectrum is the same everwhere in moduli space (and that it only consists of BPS states with $j^+=j^-$). 

We should also stress that the moduli correspond to special BPS states of this kind:  they are described by the states with 
$j^+=j^-=\frac{1}{2}$. In particular, our dual CFT has therefore the same number of moduli (namely one complex modulus) as
supergravity or the world-sheet description. 

Finally, one might wonder whether one can compare the elliptic genus for this $\mathcal{N}=2$ subalgebra as this would also explore the $1/4$ BPS spectrum. 
However, as remarked in \cite{Gukov:2004fh}, the elliptic genus for this subalgebra always vanishes, as it did in the case of $\mathbb{T}^4$.\footnote{One might ask whether we could orbifold S$^3\times$S$^{1}$ to obtain another 4D HKT manifold that has non-vanishing index, as is the case for  $\mathbb{T}^4$ that leads to K3 upon taking a $\mathbb{Z}_2$ orbifold. However this is not possible, since there are no other 4d HKT spaces.}
It is possible to modify the index so that it does not vanish  \cite{Gukov:2004fh}. 
(This is the natural analogue of the construction of \cite{Maldacena:1999bp} for the case of $\mathbb{T}^4$.) 
However, in either case, the resulting index does
not give rise to interesting constraints. 
For example, in our context only the untwisted sector contributes to the low-lying spectrum, while the contributions from the twisted sectors cancel out. 

\subsection{An effective $\mathbb{T}^2$ description}

The form of the chiral primary spectrum is formally the same as that of the symmetric orbifold of $\mathbb{T}^2$. In order
to see this, recall that a large $\mathcal{N}=4$ BPS multiplet always contains two $\mathcal{N}=2$ chiral primaries: 
in addition to the highest weight, the state that is obtained by the action of the free fermion $Q_{-1/2}^{++}$
is also chiral primary. Thus the chiral primary spectrum is given by $\tfrac{c}{6}$ overlapping diamonds of the form
\be \label{hodge}
\begin{tabular}{ccc}
& 1 & \\
1 & & 1 \\
& 1 &
\end{tabular}
\ee
Since (\ref{hodge}) is the Hodge diamond of $\mathbb{T}^2$, the BPS spectrum has the same form as the symmetric orbifold
of $\mathbb{T}^2$. This also has a neat microscopic interpretation, at least for $\kappa=0$: the algebra $\mathcal{S}_0$ consists 
of $4$ free fermions and one boson --- the $\mathfrak{su}(2)_\kappa$ theory disappears at $\kappa=0$. 
Two fermions of the four are uncharged w.r.t.~the $\mathfrak{u}(1)$-current ($\psi^{+-}$ and $\psi^{-+}$). These two fermions only appear together as a bilinear combination in the 
generators of the $\mathcal{N}=2$ subalgebra, as one can see from \eqref{realization1} 
and \eqref{N2subalgebra}. 
We can thus bosonize the two uncharged fermions and write down the $\mathcal{N}=2$ algebra without the use of vertex operators. We thus 
obtain a torus theory, where the $\mathcal{N}=2$ algebra agrees with the canonical ${\cal N}=2$ structure on $\mathbb{T}^2$.  
Among other things, this allows us to copy well-known results for the chiral ring of the $\sigma$-model on $\mathbb{T}^2$ for $\mathcal{S}_0$. 
In particular the chiral ring of the $Q_1$-fold symmetric orbifold of $\mathcal{S}_0$ can be identified with the Dolbeault cohomology ring of $(\mathbb{T}^2)^{Q_1}/S_{Q_1}$. 
It is well-known that this cohomology ring has the structure of a Fock space of free particles \cite{Vafa:1994tf, Dijkgraaf:1996xw, Dijkgraaf:1998zd, deBoer:2008ss}, one particle for every cohomology element of $\mathbb{T}^2$. 
Furthermore, the cohomology of the $Q_1$-fold symmetric product corresponds to the subset in this Fock space with conformal weight $Q_1$. 

In particular, it is easy to write down a generating function for the Poincar\'e polynomial of the symmetric product 
\cite[eq. (5.4)]{deBoer:1998kjm}, see also \cite{Soergel:1993aa} for a mathematical treatment for the case of
$\mathrm{K3}$
\be 
\sum_{Q_1=0}^\infty p^{Q_1}P_{t,\bar{t}}(\textrm{Sym}^{Q_1}(\mathcal{S}_0))=\prod_{m=1}^\infty 
\frac{(1+p^m t^{\frac{1}{2}(m+1)}\bar{t}^{\frac{1}{2}(m-1)})(1+p^m t^{\frac{1}{2}(m-1)}\bar{t}^{\frac{1}{2}(m+1)})}{
(1-p^m t^{\frac{1}{2}(m-1)}\bar{t}^{\frac{1}{2}(m-1)})(1-p^m t^{\frac{1}{2}(m+1)}\bar{t}^{\frac{1}{2}(m+1)})}\ .
\ee
Here, the variable $m$ parametrizes the length of the cycle, and to extract the contribution from a single-twist cycle of length $n$, we 
take one term coming from $m=n$, with all the other terms coming from the vacuum contribution $m=1$ and the term $(1-p)^{-1}$. 
More specifically, the single-twist contribution from the $n$-twisted sector turns out to be 
\be 
\sum_{Q_1=n}^\infty p^{Q_1}P^{n-\text{twist}}_{t,\bar{t}}(\textrm{Sym}^{Q_1}(\mathcal{S}_0))
=\frac{p^nt^{\frac{1}{2}n}\bar{t}^{\frac{1}{2}n}}{1-p}(t^{\frac{1}{2}}\bar{t}^{\frac{1}{2}}
+t^{\frac{1}{2}}\bar{t}^{-\frac{1}{2}}+t^{-\frac{1}{2}}\bar{t}^{\frac{1}{2}}+t^{-\frac{1}{2}}\bar{t}^{-\frac{1}{2}})
\ee
and in particular does not change when increasing $Q_1$ beyond $n$. 
We see that the exponents of $t$ and $\bar{t}$ are only integers when $n$ is odd. Hence only those states can lift to BPS 
representations $[j,j,u=0]_S$ of the large $\mathcal{N}=4$ algebra, since these states always have half-integer charges, 
i.e.~integer exponents of $t$ and $\bar{t}$. The contribution is then simply given by a diamond at height $\tfrac{1}{2}(n-1)$, and 
since $n \le Q_1$, the maximal achievable exponent of $t$ (or $\bar{t}$) is $\tfrac{1}{2}Q_1$. This then reproduces
precisely the chiral primary spectrum of eq.~\eqref{BPS_spectrum Skappa} with the correct cutoff.

\section{Discussion and outlook}\label{sec:discussion}

In this paper we have identified the CFT dual of string theory on $\mathrm{AdS}_3 \times \mathrm{S}^3  \times \mathrm{S}^3 \times \mathrm{S}^1$ 
for the case that the larger of the two $Q_5^\pm$ quantum numbers is a multiple of the smaller one: for the case that $Q_5^- \geq Q_5^+$, the relevant CFT is the symmetric orbifold 
\be
\textrm{Sym}^{Q_1 Q^+_5} (\mathcal{S}_{\kappa}) \qquad \textrm{with} \quad \kappa=\frac{Q_5^-}{Q_5^+}-1\ . \label{S kappa generald}
\ee
This proposal was motived by considering a brane construction, from which the CFT dual could be read off; this argument is fairly clean provided that $Q_5^+=1$, while for $Q_5^+>1$ our proposal is more of an educated guess. In either case, however, we have managed to give convincing evidence for this proposal. 
In particular, we have shown that the BPS spectra of the dual CFT reproduces exactly that of supergravity or the world-sheet description. 
In fact, the agreement is as good as for the familiar case of $\mathbb{T}^4$. 
As a consequence the moduli also match. Finally, our proposal also has the right behaviour in the limit in which the radius of the ${\rm S}^3$ goes to infinity: the chiral algebra of the dual CFT then becomes that of the symmetric orbifold of $\mathbb{T}^4$, as expected, see also \cite{Gaberdiel:2014cha}. 

It would be very interesting to perform further tests on the proposal; for example, it would be very interesting to compare $3$-point functions as in 
\cite{Gaberdiel:2007vu,Dabholkar:2007ey}. 
It would also be very interesting to understand to which extent our proposal fits together with the UV description suggested in \cite{Tong:2014yna}, and how it relates to the large ${\cal N}=4$ superconformal higher spin -- CFT duality of \cite{Gaberdiel:2013vva}. 
We hope to come back to some of these questions in the near future.

\section*{Acknowledgements}

We are grateful to Rajesh Gopakumar for many useful discussions as well as for initial collaboration. We also thank Ofer Aharony, Costas Bachas, Kevin Ferreira, Sergei Gukov, Juan Maldacena, Emil Martinec, Greg Moore, Alessandro Sfondrini, Andy Strominger, David Tong, Edward Witten and Zhixian Zhu for 
useful conversations. MRG thanks the ITP at the Chinese Academy of Sciences for hospitality at various stages of this work. The research of LE and MRG is 
(partly) supported by the NCCR SwissMAP, funded by the Swiss National Science Foundation. All of three of us gratefully acknowledge the hospitality of the 
Galileo Galilei Institute for Theoretical Physics (GGI) for hospitality, and INFN for partial financial support during the program ``New Developments in AdS3/CFT2 Holography".

\appendix

\section{The large ${\cal N}=4$  algebra and its BPS bound}\label{app:N=4}

In bi-spinor notation, the large ${\cal N}=4$ superconformal algebra $A_\gamma$  is generated by
\begin{align}
{}[U_m,U_n]  = &  \tfrac{k^+ + k^-}{2} \, m \, \delta_{m,-n}  \label{A10} \\
{}[A^{+, i}_m, Q^{\mu\nu}_r]  = &  \tfrac{1}{2}\tensor{(\sigma^i)}{_\rho^\mu} \, Q^{\rho\nu}_{m+r}   \label{A11} \\
{}[A^{-, i}_m, Q^{\mu\nu}_r]  = &  \tfrac{1}{2}\tensor{(\sigma^i)}{_\rho^\nu} \, Q^{\mu\rho}_{m+r}   \label{A12} \\
{} \{Q^{\mu\nu}_r,Q^{\rho\tau}_s \}  = &  (k^+ + k^-) \, \epsilon^{\mu\rho}\epsilon^{\nu\tau} \, \delta_{r,-s}  \label{A13} \\
{}[A^{\pm, i}_{m}, A^{\pm, j}_{n} ]  = &  \tfrac{k^\pm}{2}\, m \, \delta^{ij}\, \delta_{m,-n} 
+ {\rm i}\, \epsilon^{ijl}\, A^{\pm, l}_{m+n}  \label{A14} \\
{} [U_m,G^{\mu\nu}_r]  = & {\rm i}\,m \, Q^{\mu\nu}_{m+r}  \label{A15} \\
{}[A^{+, i}_{m},G^{\mu\nu}_r]  = &  \tfrac{1}{2}\tensor{(\sigma^i)}{_\rho^\mu} \,G^{\rho\nu}_{m+r}  + (1-\gamma)\, m \, \tensor{(\sigma^i)}{_\rho^\mu}\, Q^{\rho\nu}_{m+r}  \label{A16} \\
{}[A^{-, i}_{m},G^{\mu\nu}_r]  = &  \tfrac{1}{2}\tensor{(\sigma^i)}{_\rho^\nu} \,G^{\mu\rho}_{m+r}  - \gamma\, m \, \tensor{(\sigma^i)}{_\rho^\nu}\, Q^{\mu\rho}_{m+r}  \label{A17} \\
{} \{Q^{\mu\nu}_r,G^{\rho\tau}_s\}  = & 2\,  \epsilon^{\mu\pi}\tensor{(\sigma_i)}{_\pi^\rho}\epsilon^{\nu\tau}\, A^{+, i}_{r+s} - 2 \epsilon^{\nu\pi}\tensor{(\sigma_i)}{_\pi^\tau}\epsilon^{\mu\rho}\, A^{-, i}_{r+s} + 2 {\rm i}\,\epsilon^{\mu\rho}\epsilon^{\nu\tau} \, U_{r+s}  \label{A18} \\
{} \{G^{\mu\nu}_r,G^{\rho\tau}_s\}  = & -\tfrac{2c}{3}\, \epsilon^{\mu\rho}\epsilon^{\nu\tau}\, (r^2 - \tfrac{1}{4}) \delta_{r,-s} 
- 4\epsilon^{\mu\rho}\epsilon^{\nu\tau}\, L_{r+s} \nonumber \\
 &  + 4\, (r-s)\, \left(\gamma\, \epsilon^{\mu\pi}\tensor{(\sigma^i)}{_\pi^\rho}\epsilon^{\nu\tau}\, A^{+, i}_{r+s} +(1-\gamma)\, \epsilon^{\nu\pi}\tensor{(\sigma^i)}{_\pi^\tau}\epsilon^{\mu\rho} A^{-, i}_{r+s} \right) \ .
 \label{A19}
\end{align}
In terms of the levels of the two $\mathfrak{su}(2)$ algebras, we have
\begin{equation}\label{A9}
\gamma = \frac{k^-}{k^+ + k^-} \ , \qquad c = \frac{6 k^+ k^-}{k^+ + k^-} \ .
\end{equation}
Here, greek indices $\mu,\nu,\dots$ indices are spinor indices and get as usual raised and lowered by the epsilon symbol $\epsilon_{\mu\nu}$, which we have indicated explicitly. Indices $i,j,\dots$ are adjoint indices of $\mathfrak{su}(2)$. Finally, $\sigma^i$ denotes the Pauli matrices, i.e.~the two-dimensional spinor representation of $\mathfrak{su}(2)$.
\subsection{The BPS Bound}\label{app:bound}

The highest weight representations of the large superconformal ${\cal N}=4$ algebra $A_\gamma$ are characterized by 
$(h,j^\pm,u)$, where $h$ is the conformal dimension of the highest weight states, while $j^\pm$ are the spins of the two affine 
$\mathfrak{su}(2)$ algebras, and $u$ denotes the $\mathfrak{u}(1)$-charge, i.e.\ the eigenvalue under $U_0$. Unitarity implies that 
$j^\pm \leq \frac{k^\pm}{2}$.  However, as explained in \cite{Gunaydin:1988re}, unitarity actually requires that 
\begin{equation}\label{A.23}
j^\pm \leq \frac{k^\pm -1}{2} \ .
\end{equation}
The BPS bound takes the form \cite{Gunaydin:1988re,Petersen:1989zz,Petersen:1989pp}
\begin{equation}\label{A28}
h \geq \frac{1}{k^++k^-} \, \Bigl[ k^+ j^- + k^- j^+ + u^2 + (j^+-j^-)^2 \Bigr] \ .
\end{equation}
Note that this bound differs from the the corresponding BPS bound of the wedge algebra 
$D(2,1|\alpha)$, whose BPS bound is  \cite{deBoer:1999rh,Gukov:2004ym}
\begin{equation}\label{A30}
h \geq \Bigl[ \frac{1}{1+\alpha}\, j^- + \frac{\alpha}{1+\alpha} \, j^+ \Bigr]  \ .
\end{equation}
Indeed, apart from the additional $u^2$ term there is in particular also the $(j^+-j^-)^2$ term. If we denote the corresponding 
representation by $[j^+,j^-,u]$ then it only satisfies the BPS bound of $D(2,1|\alpha)$ if $u=0$ and $j^+=j^-$. On the other hand, if 
this is the case, the BPS representation $[j^+,j^-,u]$ of the linear $A_\gamma$ algebra contains actually two BPS 
representations of $D(2,1|\alpha)$ 
\be\label{largeN4BPS}
[j,j,u=0]_S  = [j,j]_s \oplus [j +\tfrac{1}{2},j \oplus \tfrac{1}{2}]_s \oplus \ \hbox{non-BPS reps of $D(2,1|\alpha)$}\ .
\ee
This is basically a consequence of the fact that in addition to the four supercharges (that also appear in 
$D(2,1|\alpha)$), $A_\gamma$ also contains four free fermions. In particular, every BPS representation of 
$A_\gamma$ contains always two BPS states whose spins differ by ${j^\pm}' = j^\pm + \frac{1}{2}$.

\subsection{The $\mathcal{N}=2$ subalgebra} \label{app:N2subalgebra}

The large superconformal $\mathcal{N}=4$ algebra contains a superconformal $\mathcal{N}=2$ algebra \cite{Gukov:2004ym}. Set
\be 
J=2i(\gamma A^{+,3}+(1-\gamma)A^{-,3})\ , \label{N2subalgebra}
\ee
this constitutes together with $G^{++}$, $G^{--}$ and the energy-momentum tensor an $\mathcal{N}=2$ algebra. Chiral primaries of this $\mathcal{N}=2$ algebra correspond to BPS states of the large $\mathcal{N}=4$ algebra of the form $(j,j,u=0)_S$. Moreover, by \eqref{largeN4BPS}, every short representation of the form $[j,j,u=0]_S$ of the large $\mathcal{N}=4$ algebra contains two chiral primaries of the $\mathcal{N}=2$ subalgebra.

\section{Superconformal affine algebras}\label{app:saffine}

In this appendix we review briefly the structure of superconformal affine algebras. We will only be interested in two
examples, the algebra $\mathfrak{sl}(2,\mathbb{R})^{(1)}_k$ and the algebra $\mathfrak{su}(2)^{(1)}_{k'}$. For 
the former we choose a basis as 
\begin{alignat}{5}
\bigl[\JJ^{+}_{m},\JJ^{-}_{n}\bigr] 
={}&
 -2\JJ^{3}_{m+n} + km\delta_{m,-n}&
 & \ \  &
 \bigl[\JJ^{3}_{m},\JJ^{\pm}_{n}\bigr] 
={}&
 \pm \JJ^{\pm}_{m+n}&
  & \ \ &
 \bigl[\JJ^{3}_{m},\JJ^{3}_{n}\bigr] 
={}&
 -\frac{k}{2}m\delta_{m,-n}
 \nonumber\\
 \bigl[\JJ^{\pm}_{m},\psi^{3}_{r}\bigr]
={}&
 \mp\psi^{\pm}_{m+r} &
 & \ \  &
 \bigl[\JJ^{3}_{m},\psi^{\pm}_{r}\bigr]
={}&
 \pm\psi^{\pm}_{m+r}&
  &\ \  &
 \bigl[\JJ^{\pm}_{m},\psi^{\mp}_{r}\bigr]
 ={}&
 \mp 2\psi^{3}_{m+r} \nonumber  \\
 \bigl\{\psi^{+}_{r},\psi^{-}_{s}\bigr\}
={}&
 k\delta_{r,-s}&
 & \ \ &
 \bigl\{\psi^{3}_{r},\psi^{3}_{s}\bigr\}
={}&
 -\frac{k}{2}\delta_{r,-s}\ .  & \label{sl2s}
\end{alignat}
On the other hand, the affine $\mathfrak{su}(2)^{(1)}_{k'}$ generators satisfy 
\begin{alignat}{5}
\bigl[\KK^{+}_{m},\KK^{-}_{n}\bigr] 
={}&
 2\KK^{3}_{m+n} + k' m\delta_{m,-n}&
 &\ \ &
 \bigl[\KK^{3}_{m},\KK^{\pm}_{n}\bigr] 
={}&
 \pm \KK^{\pm}_{m+n}&
 &\ \ &
 \bigl[\KK^{3}_{m},\KK^{3}_{n}\bigr] 
={}&
 \frac{k'}{2}m\delta_{m,-n}
 \nonumber\\
 \bigl[\KK^{\pm}_{m},\chi^{3}_{r}\bigr]
 ={}&
 \mp\chi^{\pm}_{m+r}&
 &\ \ &
 \bigl[\KK^{3}_{m},\chi^{\pm}_{r}\bigr]
 ={}&
 \pm\chi^{\pm}_{m+r}&
 &\ \ &
 \bigl[K^{\pm}_{m},\chi^{\mp}_{r}\bigr]
 ={}&
 \pm 2\chi^{3}_{m+r}
 \nonumber \\
 \bigl\{\chi^{+}_{r},\chi^{-}_{s}\bigr\}
 ={}&
 k'\delta_{r,-s}&
 &\ \ &
 \bigl\{\chi^{3}_{r},\chi^{3}_{s}\bigr\}
 ={}&
 \frac{k'}{2}\delta_{r,-s}\ .& \label{su2s}
\end{alignat}
In each case it is possible to decouple the fermions from the bosons, i.e.~to redefine the bosonic generators by bilinears in the
fermions so that they commute with the fermions. For $\mathfrak{sl}(2,\mathbb{R})^{(1)}_k$, the resulting decoupled
bosonic algebra then has level $k+2$, while for $\mathfrak{su}(2)^{(1)}_{k'}$ the level of the decoupled bosonic
algebra is $k'-2$. 

\subsection{Spectral flow automorphism}

For the description of the spectrally flowed representations the existence of a family of automorphisms is important. For 
any $w\in\mathbb{Z}$, we define new generators as 
\be\label{spectralflow}
\begin{array}{rclrcl}
{\displaystyle \hat{\JJ}^\pm_n} & = & {\displaystyle \JJ^\pm_{n\mp w}}  \qquad & 
{\displaystyle \hat{\KK}^\pm_n} & = & {\displaystyle \KK^\pm_{n\pm w}} \\[2pt]
{\displaystyle \hat{\JJ}^3_n} & = & {\displaystyle \JJ^3_n + \tfrac{k}{2} w \delta_{n,0} } \qquad &
{\displaystyle \hat{\KK}^3_n} & = & {\displaystyle \KK^3_n + \tfrac{k'}{2} w \delta_{n,0} } \\[2pt]
{\displaystyle \hat{L}^{\mathfrak{sl}}_n}  & = & {\displaystyle L^{\mathfrak{sl}}_n - w \JJ^3_n - \tfrac{k}{4} w^2 \delta_{n,0} } \qquad \qquad &
{\displaystyle \hat{L}^{\mathfrak{su}}_n}  & = & {\displaystyle L^{\mathfrak{su}}_n + w \KK^3_n + \tfrac{k'}{4} w^2 \delta_{n,0}} \\[2pt]
{\displaystyle \hat{\psi}^3_r} & = & {\displaystyle \psi^3_r } \qquad  
& {\displaystyle \hat{\chi}^3_r} & = & {\displaystyle \chi^3_r }  \\[2pt]
{\displaystyle \hat{\psi}^{\pm}_r} & = & {\displaystyle \psi^{\pm}_{r\mp w}} \qquad & 
{\displaystyle \hat{\chi}^{\pm}_r} & = & {\displaystyle \chi^{\pm}_{r\pm w} \ .}
\end{array}
\ee
One verifies easily that these new generators also satisfy the commutation relations of the superconformal affine algebra, 
i.e.~eqs.~(\ref{sl2s}) and (\ref{su2s}) above. In addition, they satisfy the relations
\be
{} [L_m^{\mathfrak{sl}}, \JJ^a_n] = - n \, \JJ^a_{m+n} \ , \qquad
[L_m^{\mathfrak{su}}, \KK^a_n] = - n  \, \KK^a_{m+n}  \ , 
\ee
and similarly for the fermions, 
\be
{} [L_m^{\mathfrak{sl}}, \psi^a_n] = \bigl(- \tfrac{m}{2} - n)  \, \psi^a_{m+n} \ , \qquad
[L_m^{\mathfrak{su}}, \chi^a_n] = \bigl( - \tfrac{m}{2} - n) \, \chi^a_{m+n}  \ .
\ee

\section{$\mathcal{S}_\kappa$ theory and its chiral algebra} \label{subsec:Skappa algebra}

Since the $\mathcal{S}_\kappa$ algebra, i.e.~the chiral algebra of the $\mathcal{S}_\kappa$ theory, will play a central role for the remainder of the paper, we shall briefly review its structure here. 
The superconformal affine algebra $\mathfrak{su}(2)^{(1)}_{\kappa+2}$ is generated by bosonic generators (that define an affine $\mathfrak{su}(2)$ algebra at level $\kappa+2$), as well as free fermions in the adjoint representation of $\mathfrak{su}(2)$,
see Appendix~\ref{app:saffine} for an explicit description. 
As is also mentioned there, it is possible to decouple the fermions from the bosons, and the resulting (decoupled) generators (that we shall denote by $J^i$ with $i=1,2,3$) then have level $\kappa$. 
The $\mathfrak{u}(1)^{(1)}$ algebra, on the other hand, consists of a single free boson that we shall denote by $\partial\phi$, as well as a single free fermion. 
(For $\mathfrak{u}(1)$, there is no need to decouple the fermion since the commutators in the adjoint representation vanish anyway.) 
Together with the three fermions from $\mathfrak{su}(2)^{(1)}$ we therefore have altogether four decoupled free fermions that we denote by $\psi^{\mu\nu}$ with $\mu,\nu =1,2$ being bispinor indices as explained in Appendix~\ref{app:N=4}. 
The commutation relations of the associated modes are then 
\begin{equation}
\begin{aligned}
{}[J^i_n,J^j_m] & =  \epsilon_{ijl} J^l_{m+n} + \kappa\, m\, \delta^{ij} \, \delta_{m,-n} \\
{}[\alpha_m, \alpha_n] & =   m \, \delta_{m,-n}  \label{JJcom}\\
\{\psi^{\mu\nu}_r,\psi^{\rho\tau}_s\} & =  \epsilon^{\mu\rho}\epsilon^{\nu\tau} \delta_{r,-s} \ ,
\end{aligned}
\end{equation}
where we have denoted the modes of the free boson $\partial\phi$ by $\alpha_m$.
The bilinears in the fermions generate the current algebra $\mathfrak{so}(4)_1 \cong \mathfrak{su}(2)_1 \oplus \mathfrak{su}(2)_1$,
with respect to which they transform in the representation $\mathbf{4}=(\mathbf{2},\mathbf{2})$. The $\mathcal{S}_\kappa$ algebra contains the
large ${\cal N}=4$ superconformal algebra (whose definition we spell out, for the convenience of the reader, in Appendix~\ref{app:N=4}), 
where the associated fields are defined as \cite{Sevrin:1988ew}
\begin{equation}
\begin{aligned}
L_m  = &  \frac{1}{2}(\alpha \alpha)_m + \frac{(J^i J^i)_m}{\kappa+2}-\frac{1}{4}\epsilon_{\mu\rho}\epsilon_{\nu\tau}(\psi^{\mu\nu} \partial \psi^{\rho\tau})_m  \label{realization1}\\
A^{+,i}_m  = &  \frac{1}{8}\epsilon_{\nu\tau}\tensor{(\sigma^i)}{_\mu^\pi}\epsilon_{\pi\rho}(\psi^{\mu\nu}\psi^{\rho\tau})_m  \\
A^{-,i}_m  = &  \frac{1}{8}\epsilon_{\mu\rho}\tensor{(\sigma^i)}{_\nu^\pi}\epsilon_{\pi\tau}(\psi^{\mu\nu}\psi^{\rho\tau})_m+J^i_m\\
G^{\mu\nu}_r   = &  {\rm i}\, (\alpha \psi^{\mu\nu})_r-\frac{1}{3}\sqrt{\tfrac{2}{\kappa+2}}\tensor{(\sigma_i)}{_\rho^\mu} (A^{+,i} \psi^{\rho\nu})_r 
+\frac{2}{3}\sqrt{\tfrac{2}{\kappa+2}}\tensor{(\sigma_i)}{_\rho^\nu} (J^i \psi^{\mu\rho})_r \\
&   -\frac{1}{3}\sqrt{\tfrac{2}{\kappa+2}}\tensor{(\sigma_i)}{_\rho^\nu} (A^{-,i} \psi^{\mu\rho})_r\\
U_m  = &  \sqrt{\frac{\kappa+2}{2}} \alpha_m \\
Q^{\mu\nu}_r  = &  \sqrt{\frac{\kappa+2}{2}}\psi^{\mu\nu}_r \ .
\end{aligned}
\end{equation}
In this realization, the two affine  $\mathfrak{su}(2)$ algebras appear at level 
\be\label{levels}
Q_5^+ = k^+=1 \qquad \textrm{and} \qquad Q_5^- = k^-=\kappa+1 \ . 
\ee 
The central charge of the ${\cal S}_\kappa$ theory is then
\be\label{centSk}
c(\mathcal{S}_\kappa) = \frac{3 \kappa}{\kappa+2} + 3 = 6\, \frac{\kappa+1}{\kappa+2} \ . 
\ee

It is obvious that the symmetric orbifold \eqref{S kappa general}  inherits the large ${\cal N}=4$ superconformal symmetry from its seed theory, 
and that its levels are $(Q_1Q_5^+,Q_1Q_5^-)$. It therefore has the correct charges to match the expectations from \cite{Gukov:2004ym, Elitzur:1998mm}. 

\section{Character derivation of the twisted sector spectrum} \label{app:character}
In this appendix, we derive the claims made about twist sectors of the symmetric product orbifold of $\mathcal{S}_\kappa$ in the main text. 
The partition function of a single $\mathcal{S}_\kappa$-theory reads
\begin{align}
Z(q,y,z)&=Z_\text{bos}(q)Z_{\mathfrak{su}(2)_\kappa}(q,z) \Big|q^{-\frac{1}{12}}\prod_{m=1}^\infty(1+y^{\frac{1}{2}}z^{\frac{1}{2}}q^{m-\frac{1}{2}})(1+y^{-\frac{1}{2}}z^{\frac{1}{2}}q^{m-\frac{1}{2}})\nonumber\\
&\qquad\qquad\qquad\qquad\qquad\times(1+y^{\frac{1}{2}}z^{-\frac{1}{2}}q^{m-\frac{1}{2}})(1+y^{-\frac{1}{2}}z^{-\frac{1}{2}}q^{m-\frac{1}{2}})\Big|^2\\
&=Z_\text{bos}(q)Z_{\mathfrak{su}(2)_\kappa}(q,z)\abs{\frac{\vartheta_3(\frac{1}{2}(\xi+\zeta);\tau)\vartheta_3(\frac{1}{2}(\xi-\zeta);\tau)}{\eta(\tau)^2}}^2\ .
\end{align}
Here, $Z_\text{bos}(q)$ and $Z_{\mathfrak{su}(2)_\kappa}$ are the partition functions of the free boson and of $\mathfrak{su}(2)_\kappa$,
respectively. We shall not need their precise forms, only the fact that they are modular invariant. Finally, 
$y=e^{2\pi i\xi}$ and $z=e^{2\pi i\zeta}$ are the  chemical potentials for the two $\mathfrak{su}(2)$'s.

\subsection{Odd twist}
Consider an odd cyclic twist of length $n$. Then
\be 
\text{\scalebox{.7}{$1$}}\underset{(1\cdots n)}{\raisebox{-5pt}{\text{\scalebox{2}{$\square$}}}}=Z(q^n,y^n,z^n)\, \bigl(Z(q,y,z)\bigr)^{N-n} \ .
\ee
We now perform an S-modular transformation to relate this to the sector where the boundary conditions along the two cycles of the
torus are interchanged. Omitting phase factors obtained from the Jacobi form transformations, this leads to 
\be 
\text{\scalebox{.7}{$(1\cdots n)$}}\underset{1}{\raisebox{-5pt}{\text{\scalebox{2}{$\square$}}}}=Z(q^{\frac{1}{n}},y,z)\, \bigl(Z(q,y,z)\bigr)^{N-n} \ , \label{partition_function_twisted_sector_odd_twist}
\ee
where we have used the modular properties of the theta functions. 
Thus, we can interpret the states as generated by the usual operators, but fractionally moded. 
Furthermore, in order to determine the ground state energy we note that the leading term is $q^{-\frac{c}{24n} - \frac{c}{24} (N-n)}$, 
where $c$ is the central charge of $\mathcal{S}_\kappa$. Since the total central charge of the symmetric orbifold is $cN$, the 
ground state energy relative to the vacuum is
\be 
h=\frac{c}{24n}(n^2-1)\ .
\ee
Furthermore, since  in \eqref{partition_function_twisted_sector_odd_twist}, $q^{\frac{1}{n}}$ instead of $q$ appears, the 
conformal weights are divided by a factor $n$. This then yields eq.~(\ref{twisted_sector_state_odd_eq}).

\subsection{Even twist}
For even twist the story is more subtle. Combining an even number of fermions changes the statistics \cite{Lunin:2001pw}, so the character has an additional $(-1)^F$ inserted. Thus, in this case
\be 
\text{\scalebox{.7}{$1$}}\underset{(1\cdots n)}{\raisebox{-5pt}{\text{\scalebox{2}{$\square$}}}}=Z_\text{bos}(q^n)Z_{\mathfrak{su}(2)_\kappa}(q^n,z^n)\abs{\frac{\vartheta_4(\frac{n}{2}(\xi+\zeta);n\tau)\vartheta_4(\frac{n}{2}(\xi-\zeta);n\tau)}{\eta(n\tau)^2}}^2\, \bigl(Z(q,y,z)\bigr)^{N-n} \ ,
\ee
where now different theta functions appear. Under a modular $S$-transformation, $\vartheta_4$ transforms into $\vartheta_2$, and we obtain 
\begin{align} 
\text{\scalebox{.7}{$(1\cdots n)$}}\underset{1}{\raisebox{-5pt}{\text{\scalebox{2}{$\square$}}}}&=Z_\text{bos}(q^{\frac{1}{n}})Z_{\mathfrak{su}(2)_\kappa}(q^{\frac{1}{n}},z)|y^{\frac{1}{2}}+y^{-\frac{1}{2}}+z^{\frac{1}{2}}+z^{-\frac{1}{2}}|^2(q\bar{q})^{\frac{1}{6n}}\nonumber\\
&\quad \times\Big|\prod_{m=1}^\infty(1+y^{\frac{1}{2}}z^{\frac{1}{2}}q^{\frac{m}{n}})(1+y^{\frac{1}{2}}z^{-\frac{1}{2}}q^{\frac{m}{n}})(1+y^{-\frac{1}{2}}z^{\frac{1}{2}}q^{\frac{m}{n}})(1+y^{-\frac{1}{2}}z^{-\frac{1}{2}}q^{\frac{m}{n}})\Big|^2 \nonumber  \\
&\quad \times  \bigl(Z(q,y,z)\bigr)^{N-n} \ . 
\end{align}
Thus, again the operators are fractionally moded. For the bosons, the analysis is unchanged relative to odd twist, whereas for the fermions,
instead of the $q^{-\frac{1}{12 n}}$ we now have $q^{\frac{1}{6n}}$. Thus the ground state energy is in this case
\be 
c=\frac{cn}{24} + \frac{1}{24 n} \Bigl[ 4 - 1 - \frac{3\kappa}{\kappa+2} \Bigr] = \frac{cn}{24} +\frac{1}{4n(\kappa+2)}\ . 
\ee
Note that the last term is positive and goes away in the limit $\kappa\to \infty$, as is well-known for the case of $\mathbb{T}^4$. Since
$\vartheta_2$ appears in the twisted sector, the relevant representation has fermionic zero modes, and the ground states transform 
as $(\mathbf{2},\mathbf{1}) \oplus (\mathbf{1},\mathbf{2})$, i.e.~as the spinor representation of $\mathfrak{so}(4)$. The orbifold
projection eliminates some of the ground states since the fermionic zero-modes are odd under the orbifold action. Combining
left- and right-movers, the surviving states are then  
$\bigl[(\mathbf{1},\mathbf{2}) \otimes (\mathbf{1},\mathbf{2})\bigr] \oplus 
\bigl[ (\mathbf{2},\mathbf{1}) \otimes (\mathbf{2},\mathbf{1})\bigr]$. (Since the orbifold acts symmetrically on left- and 
right-movers, the same should be the case for the orbifold projection in the twisted sectors.)

\section{BPS states  in the spectrally flowed NS sectors} \label{app:BPS states}


In this appendix we will construct the BPS states, solving (\ref{NS_sector_massshell_condition}). 
As we explained above, the BPS states must come in pairs, i.e.~to each BPS state in the R-sector, there must be a corresponding BPS state 
in the NS sector with spins shifted by $\pm \tfrac{1}{2}$. Starting from a R-sector BPS state of subsection~\ref{app:R-sector} characterized by 
$j^\mathrm{R}$, ${j^\mathrm{R}}^\pm$, $j_0$, $j_0^\pm$, $w$ and $w^\pm$, we can write down two canonical candidates for BPS states 
in the NS-sector, one with the spins shifted up by $\tfrac{1}{2}$, the other with the spins shifted down by $\tfrac{1}{2}$:
\begin{enumerate}
\item[(i)] The following state has the spins shifted down by $\tfrac{1}{2}$:
\be 
j^\mathrm{NS}=j^\mathrm{R}-\frac{1}{2}\ , \quad {j^\mathrm{NS}}^\pm={j^\mathrm{R}}^\pm-\frac{1}{2}\ .
\ee
We can achieve this by setting $j_0$, $j_0^\pm$, $w$ and $w^\pm$ to the same values as in the R-sector.
Furthermore, we set $\delta=1$, $\delta^+=\delta^-=0$. The state has then the required quantum numbers. The mass shell condition 
\eqref{NS_sector_massshell_condition} is satisfied, provided that
\be 
w=w^++w^-\ . \label{w_case1}
\ee
Since we (implicitly) apply $w+w^++w^-=2w$ fermions due to spectral flow and since 
$\delta=1$, $\delta^\pm=0$, we have an odd number of fermions altogether (as required by the GSO projection). 
\item[(ii)] Similarly, we construct a state with spins shifted up by $\tfrac{1}{2}$:
\be 
j^\mathrm{NS}=j^\mathrm{R}+\frac{1}{2}\ , \quad {j^\mathrm{NS}}^\pm={j^\mathrm{R}}^\pm+\frac{1}{2}\ .
\ee 
Again, $j_0$, $j_0^\pm$, $w$ and $w^\pm$ take the same values as in the R-sector, but in this case we set $\delta=0$, $\delta^+=\delta^-=1$. Then the mass shell condition is satisfied provided that
\be 
w=w^++w^-+1\ . \label{w_case2}
\ee
Again, the total number of fermions is odd (as must be the case in order to satisfy the GSO projection): there is 
an odd number of fermions because of the spectral flow ($w+w^+ +w^-$ is odd), whereas now $\delta+\delta^++\delta^-=2$. 
\end{enumerate}
We will explain below that for each R-sector BPS state either $w=w^++w^-$ or $w=w^++w^-+1$ holds. As a consequence there is 
precisely one corresponding BPS state in the NS-sector to each BPS state in the R-sector. 
Depending on the arithmetic properties of the BPS state in the R-sector, \eqref{w_case1} or \eqref{w_case2} applies, which determines 
whether the R-sector BPS state is the highest weight state in the multiplet or the descendant. 

We have performed an extensive search on the computer to confirm that there are no other BPS states in the NS-sector than those 
we have constructed in this manner. (This is, of course, required by supersymmetry.)
However, in doing this analysis one has to be very careful since our description above is 
redundant. While the spectral flow in the $\mathfrak{sl}(2,\mathbb{R})$ algebra leads to genuinely new representations, spectral
flow in the two $\mathfrak{su}(2)$ algebras is just a convenient method to describe descendant states, and there are identifications.
For example, the state $(j_0^+ = \frac{k^{+}}{2}-1, \delta^{+}=1, w^{+}=0$) is equivalent to $(j_0^+=0,\delta^{+}=0, w^{+}=1)$,
and similarly for $j_0^-$. 
We can fix this ambiguity by choosing the $\delta$ variables to be either $(\delta,\delta^+,\delta^-)=(1,0,0)$ or $(\delta,\delta^+,\delta^-)=(0,1,1)$.

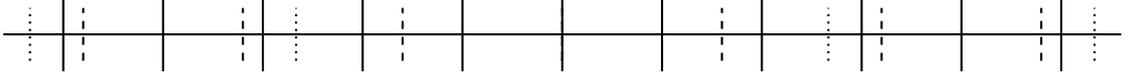
\begin{figure}
\begin{center}
\begin{tikzpicture}[scale=.7]
\draw[thick] (-10.5,0) -- (10.5,0);
\foreach \i in {-9.375,-7.5,...,9.375}
{
\draw[thick] (\i,-0.7) -- (\i,0.7);
}
\foreach \i in {-9,-6,...,9}
{
\draw[thick, dashed] (\i,-0.5) -- (\i,0.5);
}
\foreach \i in {-10,-5,...,10}
{
\draw[thick, dotted] (\i,-0.5) -- (\i,0.5);
}
\end{tikzpicture}
\end{center}
\caption{ Distribution of states in the intervals \eqref{intervals} for the case of $k^+=5$, $k^-=3$. Solid markers indicate the boundary of an interval, dotted 
markers are multiples of $\tfrac{1}{2}k^+$, dashed markers are multiples of $\tfrac{1}{2}k^-$. One can clearly see 
that there is precisely one point where all markers meet and the following solid interval does not contain any other marker. 
Apart from this exception, every two solid markers enclose exactly one dotted or dashed marker. Hence either no marker 
meets or all three markers meet at one point.} \label{distribution}
\end{figure}

To determine whether the corresponding NS-sector state has the spin shifted up or down by $\tfrac{1}{2}$, we make the following observation for the R-sector BPS states. 
Consider the following intervals for $j$
\be 
(\tfrac{1}{2}k,k]\ , \quad (k,\tfrac{3}{2}k]\ , \quad (\tfrac{3}{2}k,2k]\ , \dots \label{intervals}
\ee
corresponding to $w=1$, $2$, $3$, $\dots$. 
Then each interval contains precisely one element of the set $\tfrac{k^+}{2}\mathbb{Z}_{>0} \cup \tfrac{k^-}{2}\mathbb{Z}_{>0}$ --- with one exception. 
If the corresponding state in the set lies on the right edge of the interval, then the successive interval contains no state of the set. 
The exceptional case occurs if the spin is a multiple of $\tfrac{1}{2}\,\mathrm{lcm}(k^+,k^-)$, i.e.~if it is a multiple of both $\tfrac{k^+}{2}$ and $\tfrac{k^-}{2}$. 
This is illustrated in figure \ref{distribution} for the case of $k^+=5$, $k^-=3$.

In each fixed interval \eqref{intervals}, to the left of the dotted or dashed marker we have $w=w^++w^-+1$, while to the right of it we have 
$w=w^++w^-$ --- this just follows from the fact that the markers indicate exactly where $w^+$ (dotted), $w^-$ (dashed) or $w$ (solid) changes. 
So in between two solid markers, the spins in the NS-sector are shifted up by $\tfrac{1}{2}$ below the point where the dotted or dashed 
marker occurs, while above that point they are shifted down. Furthermore, at the position of the dotted or dashed marker itself, both solutions exist, 
so the states whose $j$ is a multiple of $\tfrac{k^+}{2}$ or $\tfrac{k^-}{2}$ occur twice in the NS-sector. The only exception occurs if a dotted or 
dasher marker coincides with a solid one --- then the two solutions have different values for $j$. 

Similarly, below a solid marker, the spins are shifted down, above they are shifted up. So when a solid marker occurs, there 
are two BPS states missing in the NS-sector. Finally consider the case when all three markers coincide. Below this triple point, 
spins are shifted up, above they are shifted down, since all of $w$, $w^+$ and $w^-$ change by one unit. Hence in this case there are 
two NS-sector BPS states sitting at this triple point, and there is no gap occurring as at the other solid markers. In summary, the NS-sector BPS spectrum therefore reads
\begin{align} 
&\bigoplus_{j \in \frac{1}{2}\mathbb{Z}_{\ge 0}\setminus\left(\frac{1}{2}\lfloor k \mathbb{Z}_{\ge 0}\rfloor \setminus \frac{1}{2}\,\mathrm{lcm}(k^+,k^-)\mathbb{Z}_{\ge 0}\, \cup \, \left(\frac{1}{2}\lfloor k \mathbb{Z}_{\ge 0}\rfloor+\frac{1}{2}\right) \setminus\left(\frac{1}{2}\,\mathrm{lcm}(k^+,k^-)\mathbb{Z}_{\ge 0}+ \frac{1}{2}\right)\right)} (j,j,u=0)_S \nonumber\\
&\qquad\qquad\qquad\qquad\ \; \oplus\ \bigoplus_{j \in \frac{k^+}{2}\mathbb{Z}_{> 0} \cup \frac{k^-}{2}\mathbb{Z}_{> 0}} (j,j,u=0)_S\ . \label{BPS_spectrum_NS_sectorapp}
\end{align}
Strictly speaking, this formula is only true for $k \ge 2$; for small values of $k^\pm$ there are some 
subtleties in the notation since then the intersection
\be 
\frac{1}{2}\lfloor k \mathbb{Z}_{\ge 0}\rfloor \setminus \frac{1}{2}\,\mathrm{lcm}(k^+,k^-)\mathbb{Z}_{\ge 0}\, \cap \, \left(\frac{1}{2}\lfloor k \mathbb{Z}_{\ge 0}\rfloor+\frac{1}{2}\right) \setminus\left(\frac{1}{2}\,\mathrm{lcm}(k^+,k^-)\mathbb{Z}_{\ge 0}+ \frac{1}{2}\right)
\ee
may be non-empty and we have to remove these states. Morally speaking, they are then removed twice from the set of the first direct sum in \eqref{BPS_spectrum_NS_sectorapp}. 

\section{Non-renormalization of the chiral ring} \label{app:chiral}

In this appendix, we show that the conformal weights of $\mathcal{N}=2$ chiral primaries is stable under deformations of the theory. 
This is probably well-known, but we could not find a good reference in the literature and therefore include a short proof 
(in the spirit of \cite{Dixon:1987bg, deBoer:2008ss, Baggio:2012rr}) for the convenience of the reader. 
In order to determine the change of conformal dimension, we need to calculate the 
two-point function of a chiral primary $\Phi_i^+$ and an anti-chiral primary $\Phi_j^-$ 
\be 
\langle \Phi^+_i(z_1)\Phi_j^-(z_2)\rangle \label{PhiPhicorrelator}
\ee
in conformal perturbation theory.\footnote{Here we have suppressed the anti-holomorphic part from our notation.}
These are the only relevant two-point functions, since the two-point function of two chiral primaries vanishes 
by conservation of the $\mathfrak{u}(1)$-charge. By the (anti-)chirality condition, $\Phi_i^+$ and $\Phi_j^-$ must 
have equal conformal weights. First order conformal perturbation theory involves the correlation functions
\be 
\langle(G_{-1/2}^+ \cdot\Phi_{1/2}^-)(z)  \Phi^+_i(z_1)\Phi_j^-(z_2)\rangle\ , \label{GPhicorrelator}
\ee
where $\Phi_{1/2}^-$ is chiral primary with conformal weight $\tfrac{1}{2}$ (the index $\tfrac{1}{2}$ is not a mode number). Similarly, also terms with $+$ and $-$ interchanged are involved. However symmetry in chiral and anti-chiral components lets us restrict to this case. We now show that \eqref{GPhicorrelator} vanishes. Writing
\be 
(G_{-1/2}^+\cdot \Phi_{1/2}^-)(z)=\oint_{w=z} \frac{\mathrm{d}w}{2\pi i}\frac{w-z_2}{z-z_2} G^+(w)\Phi^-(z) \label{GPhi}
\ee
and inserting this into the correlator \eqref{GPhicorrelator}, we can deform the contour such that it encircles $z_1$, $z_2$ and
potentially infinity (with the opposite orientation). Since $G^+(w)\Phi_i^+(z_1)$ is regular, the contour integral around $z_1$ 
vanishes. Furthermore, we 
chose the factor in the integrand of \eqref{GPhi} in such a way that the single pole at $z_2$ exactly cancels. Finally, there is 
no contribution from infinity since 
\begin{align}
\oint_{w=z} \frac{dw}{2\pi i} \frac{w-z_2}{z-z_2} G^+(w)
 &= \oint_{w=z} \frac{dw}{2\pi i} \frac{w-z_2}{z-z_2} \sum_r G^+_r w^{-r-3/2}\\
 &= \frac{1}{z - z_2} G^+_{1/2} -  \frac{z_2}{z - z_2} G^+_{-1/2}\ ,
\end{align}
and hence, inside \eqref{GPhicorrelator}, all the modes of $G^+_{-1/2}$ annihilate the in-vacuum $\langle 0|$ at infinity. 
Thus, \eqref{GPhicorrelator} vanishes.

This shows that the derivative of \eqref{PhiPhicorrelator} vanishes at every smooth point in the moduli space and thus the conformal weights of the chiral primaries are protected throughout the moduli space.

\bibliographystyle{JHEP}

\end{document}